\documentclass[aps,twocolumn,prd,nofootinbib,superscriptaddress,10pt,longbibliography,floatfix]{revtex4-2}
\usepackage{graphicx}
\usepackage{dcolumn}
\usepackage{bm}
\usepackage{hyperref}
\usepackage{lipsum}
\usepackage{xcolor}
\usepackage{calc}
\usepackage{comment}
\usepackage{float}
\usepackage{diagbox}
\usepackage{soul}
\usepackage{amsmath}
\usepackage{multirow}
\usepackage[section]{placeins}

\usepackage{ragged2e}
\usepackage{blindtext}
\usepackage{geometry}
\usepackage{appendix}
\usepackage[skip=10pt plus1pt, indent=10pt]{parskip}
\usepackage{indentfirst}
\usepackage{csquotes}
\usepackage{empheq}
\newcommand\mathcomma{\,,}

\def\dd{\mathrm{d}}
\usepackage[capitalise]{cleveref}
\usepackage[normalem]{ulem}

\DeclareUnicodeCharacter{2212}{-}

\usepackage{slashed}

\hypersetup{
    colorlinks=true,
    linkcolor=cyan,      
    urlcolor=teal,
    citecolor=blue
    }

\begin{abstract}
We update constraints on a simple model of self-interacting neutrinos involving a heavy scalar mediator with universal flavor coupling. According to past literature, such a model is allowed by Cosmic Microwave Background (CMB) data, with some CMB and large-scale structure data even favoring a strongly-interacting neutrino (SI$\nu$) scenario over $\Lambda$CDM. In this work, we re-evaluate the constraints on this model in light of the new {\it Planck} NPIPE data, DESI BAO data, and the Effective Field Theory of Large Scale Structures (EFTofLSS) applied to BOSS data. We find that {\it Planck} NPIPE are more permissive to the SI$\nu$ scenario and that DESI data favor the SI$\nu$ over $\Lambda$CDM. However, when considering EFTofBOSS data, this mode is no longer preferred. Therefore, new DESI data analyzed under the EFTofLSS are particularly awaited to shed light on this disagreement.
\end{abstract}
\begin{document}
\title{Self-interacting neutrinos in light of recent CMB and LSS data}
\author{Adèle Poudou}\email{adele.poudou@umontpellier.fr}
\author{Théo Simon}
\author{Thomas Montandon}
\author{Elsa M. Teixeira}
\author{Vivian Poulin}
\affiliation{Laboratoire Univers \& Particules de Montpellier (LUPM), CNRS \& Universit\'e de Montpellier (UMR-5299),Place Eug\`ene Bataillon, F-34095 Montpellier Cedex 05, France}

\preprint{APS/123-QED}
\renewcommand{\arraystretch}{1.5}
\newcommand{\logGeff}{$\log_{10}{G_{\rm eff, \nu}}$ }
\maketitle
\section*{Introduction}

Cosmology has entered an era of high-precision measurements, providing remarkable insights into the structure and evolution of the Universe. The $\Lambda$ Cold Dark Matter ($\Lambda$CDM) model has emerged as the standard paradigm, successfully describing observations from the Cosmic Microwave Background (CMB) to the distribution of galaxies in the late Universe. However, the fundamental nature of dark matter, dark energy, and the mechanism providing the seeds for structure formation (known as inflation) remains elusive.

In that context, recent observations have revealed two significant discrepancies between predictions of the $\Lambda$CDM model calibrated on CMB data and measurements with various probes of the Universe at low-redshift, known as the Hubble tension and $S_8$ tension. If they withstand increasingly stringent tests for systematic errors in measurements, these tensions open new avenues to learn about the dark sector. 

First and foremost, measurements using the local distance ladders consistently yield a higher value of the Hubble constant $H_0$ compared to inferences from the early Universe that rely on the {\it sound horizon}. In particular, the $H_0$ value deduced from measurements of Cepheid-calibrated Type Ia supernovae by the S$H_0$ES collaboration \cite{Riess:2021jrx} is in more than $5\sigma$ tension with that predicted by $\Lambda$CDM when fit to {\it Planck} data \cite{Planck:2018vyg}. The most promising models, in that context, appear to be those reducing the value of the sound horizon $r_s$ in the early Universe, onto which CMB and baryonic acoustic oscillations (BAO) are calibrated to infer a value of $H_0$ \cite{Schoneberg:2021qvd,Poulin:2024ken}.

Second, the $S_8$ parameter, which characterizes the amplitude of matter fluctuations on scales of $8h^{-1}$ Mpc, where $h=H_0/100$, is found smaller at the $2-3\sigma$ level by weak gravitational lensing observations \cite{Heymans:2020gsg,DES:2021wwk} than that deduced from a $\Lambda$CDM fit to CMB data, suggesting potential inconsistencies in our understanding of cosmic structure formation.
For a review of the current status of measurements and the models suggested to resolve cosmic tensions, we refer to Refs.~\cite{Abdalla:2022yfr,Verde:2023lmm}.

One of the main challenges for the models that have been proposed so far is that it is very hard to simultaneously explain a high value for $H_0$ and a low value for $S_8$. This is because our variety of cosmological observations at low-redshift provide strong constraints to the {\it shape} of the expansion history, which under $\Lambda$CDM effectively amounts to a strong constraint on the {\it fractional} matter density $\Omega_m$. As such, the increase in $H_0$ suggested by S$H_0$ES necessarily implies a significant increase in the {\it physical} matter density $\omega_{\rm m}\equiv \Omega_m h^2$,
leading to earlier matter domination, more time for structures to form and thus a {\it higher} value of $S_8$ \cite{Jedamzik:2020krr,Poulin:2024ken}. Hence, models where the $H_0$ tension is resolved often have larger $S_8$ tension or create a new tension with probes of the shape of the expansion history at late times.

In that context, the existence of exotic properties of neutrinos appears to be a promising possibility for explaining cosmic tensions. 
Nearly massless and weakly interacting particles, neutrinos represent nearly $40\%$ of the radiation budget and thus play a significant role in both the thermal history of the Universe and the growth of cosmic structures \cite{Lesgourgues:2014zoa}. In the standard cosmological model, neutrinos are treated as free-streaming particles after decoupling from the primordial plasma. However, theories involving {\it self-interaction} among neutrinos can modify their propagation and alter the way in which they affect CMB photons. 
Furthermore, the presence of additional energy density in the neutrino background provides a simple mechanism to reduce the sound horizon, while the existence of a non-zero neutrino mass affecting structure formation on small scales could explain a lower amplitude of fluctuations. 

Self-interacting neutrino models have been studied in the past as a way to alleviate the $H_0$ and $S_8$ tensions. In a model where neutrinos interact with each other, the onset of neutrino free-streaming is delayed, affecting the phase of the oscillation in the baryon-photon plasma, also known as the `neutrino drag' effect that is imprinted by free-streaming species \cite{Bashinsky:2003tk}. As a result, the cosmic microwave background (CMB) temperature and polarization power spectra shift to larger multipoles $\ell \approx \pi / \theta$, where $\theta$ represents {\it observed} angular scales, requiring an increase in the value of the Hubble constant $H_0$ to be kept fixed. In addition, the suppression of the only source of anisotropic stress in the standard model impacts the matter power spectrum by suppressing its small-scale amplitude, hereby decreasing the cosmological parameter $S_8$ \cite{Konoplich:1988mj,Berkov:1988sd,Belotsky:2001fb,Cyr-Racine:2013jua,Archidiacono:2013dua,Lancaster:2017ksf,Oldengott:2017fhy,Choi:2018gho,Song:2018zyl,Lorenz:2018fzb,Barenboim:2019tux,Forastieri:2019cuf,Smirnov:2019cae,Escudero:2019gvw,Ghosh:2019tab,Funcke:2019grs,Sakstein:2019fmf,Mazumdar:2019tbm,Blinov:2020hmc,deGouvea:2019qaz,Froustey:2020mcq,Babu:2019iml,Kreisch:2019yzn,Park:2019ibn,Deppisch:2020sqh,Kelly:2020pcy,EscuderoAbenza:2020cmq,He:2020zns,Ding:2020yen,Berbig:2020wve,Gogoi:2020qif,Barenboim:2020dmg,Das:2020xke,Mazumdar:2020ibx,Brinckmann:2020bcn,Kelly:2020aks,Esteban:2021ozz,Arias-Aragon:2020qip,Du:2021idh,CarrilloGonzalez:2020oac,Huang:2021dba,Sung:2021swd,Escudero:2021rfi,RoyChoudhury:2020dmd,Carpio:2021jhu,Orlofsky:2021mmy,Green:2021gdc,Esteban:2021tub,Venzor:2022hql,Taule:2022jrz,RoyChoudhury:2022rva,Loverde:2022wih,Kreisch:2022zxp,Das:2023npl,Venzor:2023aka,Sandner:2023ptm}.

Over the past decade, it has been surprisingly shown that CMB {\it temperature} data can equally accommodate two distinct ``modes'' in the posterior distribution, each corresponding to a significantly different self-interaction strength. The first mode effectively corresponds to the standard model quasi-massless neutrinos, with $\Lambda$CDM-like values for $H_0$ and $S_8$. Conversely, the other mode corresponds to a model comprising {\it four} strongly-interacting {\it massive} neutrino species, yielding a larger value for $H_0$ and a smaller value for $S_8$, thereby alleviating the tensions \cite{Bashinsky:2003tk,Cyr-Racine:2013jua,Archidiacono:2013dua,Lancaster:2017ksf,Oldengott:2017fhy,Barenboim:2019tux,Das:2020xke,Mazumdar:2019tbm,Brinckmann:2020bcn,RoyChoudhury:2020dmd,Kreisch:2019yzn,Park:2019ibn,Kreisch:2022zxp,Das:2023npl}.
Furthermore, recently, it has been found that the strongly-interacting mode is also favored when using alternative CMB data from the Atacama Cosmology Telescope (ACT) for both temperature and polarization \cite{Kreisch:2022zxp}, as well as for BOSS large-scale structures (LSS) data \cite{Camarena_2023} and Lyman-$\alpha$ data \cite{He:2023oke}.

However, the main issue lies in the fact that when including CMB {\it Planck} polarization data \cite{1807.06209}, the strongly-interacting mode is no longer favored and does not solve cosmic tensions \cite{Das:2020xke,Mazumdar:2020ibx,Brinckmann:2020bcn,RoyChoudhury:2020dmd,Kreisch:2019yzn}, even when paired with LSS data \cite{Camarena_2025}.
Consequently, there appears to be an inconsistency between temperature and polarization data regarding the constraints imposed on the strongly-interacting neutrino model.

Recently, the {\it Planck} collaboration has released a new set of CMB maps, called NPIPE \cite{Planck:2020olo}, in which a number of systematic effects have been improved and the signal-to-noise ratio at small scales has increased thanks to improvements in the processing of time-ordered data, allowing to use a larger sky fraction (about $\sim 80\%$) and resulting in $\sim 10\%$ higher precision on $\Lambda$CDM parameters~\cite{Rosenberg:2022sdy,Tristram:2023haj}. In addition, new BAO data by the DESI collaboration have been found to favor slightly different values of the product $H_0r_s$ \cite{DESI:2024mwx}, affecting constraints to extra relativistic species \cite{Allali:2024cji,Saravanan:2025cyi}.

 In this work, we update the constraints on the self-interacting neutrino model in light of those improved {\it Planck} data, modeled using the \textsc {CamSpec} likelihood \cite{Rosenberg:2022sdy}, employing both Bayesian and Frequentist methods to contrast both approaches. In addition, we re-evaluate constraints from LSS data using an alternative implementation of the Effective Field Theory of Large Scale Structure (EFTofLSS) applied to BOSS data, as implemented in the \texttt{PyBird} code \cite{DAmico:2020kxu}. We also include the eBOSS quasar data for the first time on top of the galaxy data that have already been used.
 Since the constraining power from all these data is stronger, one might expect that those data will help to firmly rule out strongly-interacting neutrinos as an alternative to standard model neutrinos. Yet, we find that NPIPE data, even with polarization data included, does not favor standard model neutrinos over strongly-interacting ones. 
 However, while the combination of LSS data also allows for a strongly-interacting neutrino mode, the value of the interaction strength inferred differs from that obtained from CMB data. As a result, the $\Lambda$CDM-like mode is favored when combining CMB and EFTofBOSS data.

Our paper is structured as follows. We first review the self-interacting neutrino model and its cosmological impact in \cref{model}. We then update constraints from {\it Planck} and DESI in \cref{sec:CMB} and from the EFTofBOSS and the combination with {\it Planck} in \cref{sec:LSS}. Our conclusions are summarized in \cref{conclusion}.

\section{Self-interacting neutrino model}
\label{model}
\subsection{Coupling constant}
Models of self-interacting neutrinos mediated by a scalar or pseudo-scalar mediator can typically be described through the following Lagrangian \cite{Oldengott:2014qra,Oldengott:2017fhy,Kreisch:2019yzn}:
\begin{equation}
    {\cal L}=g_{ij}\bar\nu_i\nu_j\varphi \mathcomma
\end{equation}
where $\nu$ is a left-handed neutrino Majorana spinor whose label $i$ and $j$ run over the mass eigenstates and $\varphi$ is a massive scalar particle of mass $m_\varphi$ whose coupling to the neutrinos is encoded in the matrix $g$. As is usual, we assume the mass of the mediator is large in front of the temperature of the bath, and the coupling is universal among neutrino eigenstates. As such, this interaction can be described through a four-fermion interaction similar to Fermi's model, for which the interaction rate $\Gamma_{\nu}$ of relativistic neutrinos reads
\begin{equation}
    \Gamma_{\nu}=aG^2_{\rm eff}T^5_{\nu} \mathcomma
\end{equation}
with $G_{\rm eff} \equiv |g|^2/m_\varphi^2$ the coupling constant,  $T_{\nu}$ the background temperature of neutrinos, and $a$ the scale factor \cite{Kreisch:2019yzn,Camarena_2023}. 
One can see that such an interaction rate predicts that the interaction will be active at early times and rapidly drops as $T_\nu$ decreases over time. If $G_{\rm eff}= G_{\rm F}$, where $G_{\rm F}\simeq1.166\times10^{−11}$ MeV$^{−2}$ is the Fermi constant, the interaction stops ($\Gamma_{\nu} \sim aH$) around $T_\nu\sim 1$ MeV, after which the neutrinos behave as a free-streaming species. However, for arbitrarily large $G_{\rm eff}$, the onset of free-streaming can be delayed. To alter the CMB, which decouples at $T \sim 1\,$eV, one can rapidly deduce that $G_{\rm eff}\sim 10^{10} G_{\rm F}\sim 10^{-1}$ MeV$^{−2}$ would be required. 

\subsection{Boltzmann equation for self-interacting massive neutrinos}

In the following, we assume neutrinos to be massive. We briefly review the derivation of the modified Boltzmann equation for the distribution function of neutrinos $f_{\nu}$
\begin{equation}
    \frac{\dd f_{\nu}}{\dd \tau}=C[f_{\nu}] \mathcomma
\end{equation}
 in the presence of the neutrino self-interaction and assuming a perturbed Friedmann-Lemaître-Robertson-Walker space-time metric defined in the Conformal Newtonian gauge as~\cite{Ma:1995ey} \begin{equation}
    \dd s^2 = -a^2\left[ \left( 1+2\psi(\boldsymbol{x},\tau \right) )\dd\tau^2 - \left( 1-2\phi(\boldsymbol{x},\tau) \right) \dd \boldsymbol{x}^2 \right] \mathcomma
    \end{equation}
where $\tau$ is the conformal time and $\psi$ and $\phi$ the Newtonian scalar potentials.

We expand $f_{\nu}$ to first order~\cite{Ma:1995ey}
\begin{equation}
    f_{\nu}(\boldsymbol{x},\boldsymbol{p},\tau)=f_{\nu}^{(0)}(\boldsymbol{p},\tau)[1+\Psi(\boldsymbol{x},\boldsymbol{p},\tau)] \mathcomma
\end{equation}
 assuming small perturbations $\Psi(\boldsymbol{x},\boldsymbol{p},\tau)$ around a Fermi-Dirac background distribution function $f_{\nu}^{(0)}(\boldsymbol{p},\tau)=(e^{\frac{p}{T_{\nu}}}+1)^{-1}$ (\textit{i.e.}, interactions are assumed not to modify the background distribution). The temperature fluctuations in Fourier space $\Xi_{\nu}(\boldsymbol{k},\boldsymbol{p},\tau)$ can thus be expressed as
\begin{equation}
   \Xi_{\nu}(\boldsymbol{k},\boldsymbol{p},t)= \frac{-4\Psi(\boldsymbol{k},\boldsymbol{p},t)}{\frac{\dd \ln{f_{\nu}^{(0)}}}{\dd \ln{p}}} \mathcomma
\end{equation} 
where $p \equiv |\boldsymbol{p}|$. We expand the angular dependence in a Legendre polynomial series in Fourier space
\begin{equation}
    \Xi_{\nu}(\boldsymbol{k},\boldsymbol{p},t)=\sum_\ell (-i)^\ell(2\ell+1)\nu_\ell(k,p,t)P_\ell(\mu) \mathcomma
\end{equation}
where $\mu$ is the cosine of the angle between the wave-vector $\boldsymbol{k}$ and the momentum $\boldsymbol{p}$, $P_\ell(\mu)$ are the Legendre polynomials, and $\nu_\ell(k,p,t)$ the neutrino multipole moments, which depend on the modules of the wave-vector $k$ and momentum $p$.

 The collision term $C[f_{\nu}]$ entering the Boltzmann equation can be particularly challenging to compute, even at first order, due to the momentum dependence in the temperature fluctuation $\Xi_{\nu}(\boldsymbol{x},\boldsymbol{p},\tau)$, making the full problem untractable for cosmological analyses \cite{Oldengott:2014qra,Oldengott:2017fhy,Kreisch:2019yzn}. However, it has been found that for practical purposes, neglecting the momentum dependence of the temperature fluctuations within the collision term (referred to as ``thermal approximation'' or ``relaxation time approximation'') is a reasonable approximation that allows to write the scattering term as \cite{Oldengott:2017fhy,Kreisch:2019yzn}
\begin{equation}
\begin{split}
C_{\nu}[\boldsymbol{p}]=&\frac{G^2_{\rm eff} T^6_{\nu}}{4} \frac{\partial \ln{f_{\nu}^{(0)}}}{\partial \ln{p}}\sum_\ell (-i)^\ell(2\ell+1)\nu_\ell(k,p,t)P_\ell(\mu) \\
& \times \left[ A \left(\frac{p}{T_{\nu}} \right)+ B_{\ell}\left(\frac{p}{T_{\nu}} \right) -2D_\ell \left(\frac{p}{T_{\nu}} \right)\right] \mathcomma
\end{split}
\end{equation}
where $A(x)$, $B_\ell(x)$ and $D_\ell(x)$ are functions derived in Ref.~\cite{Kreisch:2019yzn}.

Bringing it all together, the Boltzmann equations can then be written in terms of conformal time as
\begin{equation}
\begin{split}
    \frac{\partial \nu_\ell}{\partial \tau}= &-\frac{kq}{\epsilon} \left( \frac{\ell+1}{2\ell+1} \nu_{\ell+1} - \frac{\ell}{2\ell+1} \nu_{\ell-1}\right)\\
    &+4\left( \frac{\partial \phi}{\partial \tau} \delta_{\ell0}+\frac{k\epsilon}{3q} \psi\delta_{\ell1}  \right) - \frac{\Gamma_{\nu}}{f_{\nu}^{(0)}} \frac{T_{\nu,0}}{q}\\
    &\times \left[ A \left(\frac{p}{T_{\nu,0}} \right)+ B_{\ell}\left(\frac{p}{T_{\nu,0}} \right)-2D_\ell \left(\frac{p}{T_{\nu,0}} \right)\right] \nu_\ell \mathcomma
\end{split}
\end{equation}
with $T_{\nu,0}$ being the temperature of neutrinos today, $q=ap$ the comoving momentum, $\epsilon$ the total energy of neutrinos, and $\delta_{mn}$ the Kronecker delta function \cite{Kreisch:2019yzn,Camarena_2023}.
These equations have been implemented in a modified version\footnote{Our code is available at \url{https://github.com/PoulinV/class_interacting_neutrinos}.} of \texttt{CLASS} \cite{Blas:2011rf} which reproduces the implementation of Ref.~\cite{Kreisch:2019yzn} (and was used in prior works \cite{Schoneberg:2021qvd}). 

\subsection{Cosmological impact and bi-modality: reviewing the state of the art}

As previously mentioned, self-interactions between neutrinos delay the onset of their free-streaming. As free-streaming neutrinos are the only source of anisotropic stress in the standard model, self-interaction leads to particular effects on the CMB and matter power spectra.  
Of utmost importance, the  `neutrino drag' effect is a phenomenon that occurs when neutrino perturbations (and those of any free-streaming species) propagate faster than perturbations of the photon-baryon plasma, creating both a phase-shift in the photon-baryon perturbations and a reduction in their amplitude \cite{Bashinsky:2003tk,Baumann:2015rya}. 
The main observable consequence of this effect is to shift the CMB power spectrum to larger scales (smaller multipoles $\ell$) and reduce its amplitude. 
However, for sufficiently large interaction rates $\Gamma_{\nu}$, neutrino free-streaming is significantly delayed, such that the neutrino-induced phase-shift and the impact on the CMB amplitude are reduced. 
The position of the peaks in the CMB power spectra, which comes from the sum of the angular size of the sound horizon $\theta_s$ and that phase-shift, would thus be altered compared to a cosmology with weak neutrino interactions, while the tail of the CMB power spectra would experience significantly less damping \cite{Bashinsky:2003tk,Baumann:2015rya}. 
Adjusting the CMB with strongly-interacting neutrinos thus requires reshuffling cosmological parameters, opening up the possibility of resolving cosmic tensions. 

\begin{figure*}
    \centering
    \includegraphics[width=1\linewidth]{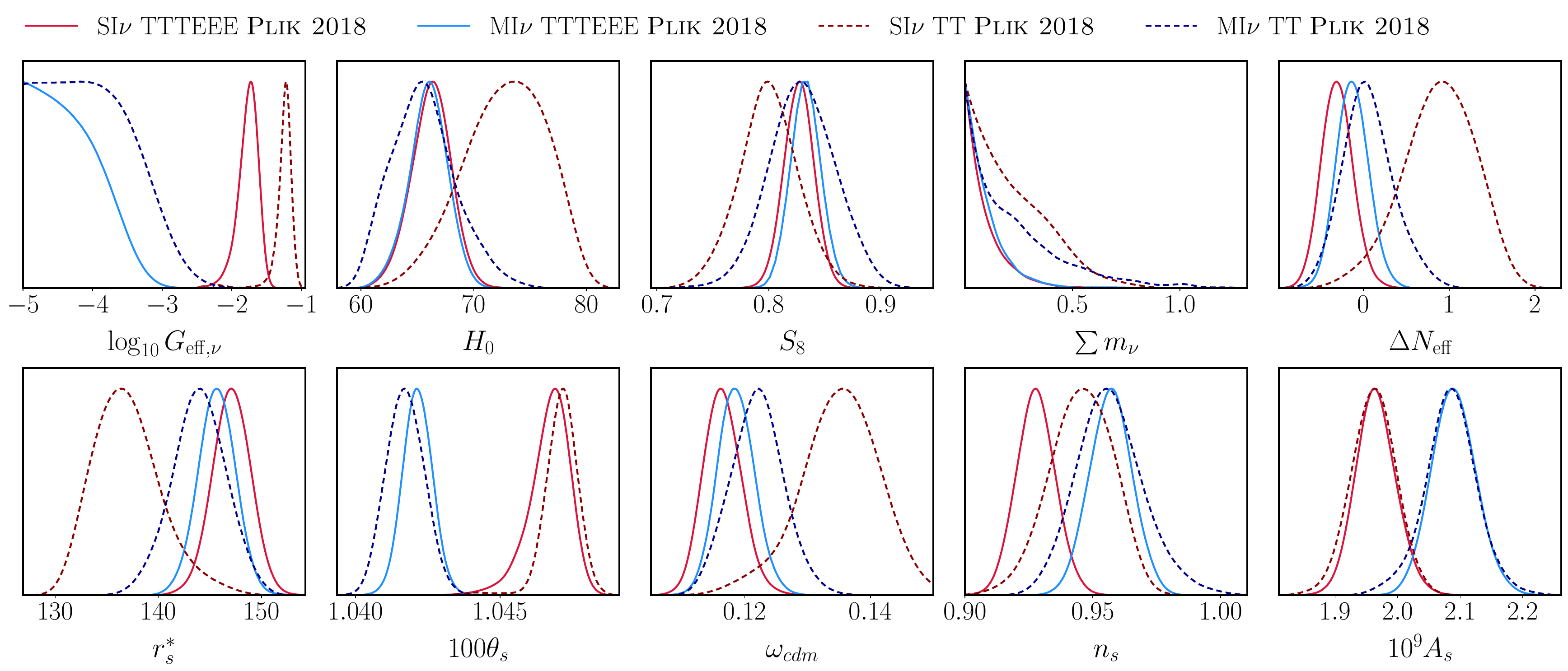}
    \caption{1-dimensional posteriors of cosmological parameters for \textsc{Plik 2018} CMB data with (filled line) and without (dashed line) polarization for ``strongly-interacting'' (red) and ``moderately interacting'' (blue) neutrinos.}
    \label{fig:TT-vs-TTTEEE}
\end{figure*}

To summarize the current status of the constraints, we present in \cref{fig:TT-vs-TTTEEE} the 1D posterior distributions of cosmological parameters of interest when analyzing the self-interacting neutrino models in light of {\it Planck} TT+low-$\ell$TT+low-$\ell$EE or {\it Planck} TTTEEE+low-$\ell$TT+low-$\ell$EE+lensing likelihoods \cite{Planck:2018vyg} using the {\sc Plik} 2018 likelihood, thereby reproducing results from the past literature \cite{Oldengott:2017fhy,Kreisch:2019yzn}. The model includes the standard $\Lambda$CDM parameters, namely the baryon and cold dark matter density $\omega_b$ and $\omega_{\rm cdm}$, the amplitude $A_s$ and tilt $n_s$ of the primordial power spectrum, the optical depth to reionization $\tau_{\rm reio}$ and the Hubble rate $H_0$, together with the three parameters describing the self-interacting neutrinos, namely $\Delta N_{\rm eff}\equiv N_{\rm eff}-3.044$, the number of relativistic species, $\sum m_\nu$ the neutrino mass sum and \logGeff~their interaction rate. Following what is common in the literature, we show results for two distinct modes: ``moderately interacting'' (MI$\nu$) neutrinos (arbitrarily defined as having $\log_{10}(G_{\rm eff,\nu}/{\rm MeV}^{-2})<2.5$) and ``strongly-interacting'' (SI$\nu$) neutrinos (having $\log_{10}(G_{\rm eff,\nu}/{\rm MeV}^{-2})>2.5$).

First, given the position of the observed peak in the posterior distribution, two cosmologies with and without neutrino free-streaming will significantly differ in the value of $\theta_s$. This is evident in \cref{fig:TT-vs-TTTEEE}, where the strongly-interacting neutrino (SI$\nu$) mode exhibits a significantly larger value of $\theta_s$ compared to the moderately interacting neutrino (MI$\nu$) mode for both data analyses.
Note that this effect alone does not necessarily lead to a substantial increase in $H_0$: while it is significant considering the size of the error bar on $\theta_s$, it remains relatively small ($\sim 0.5\%$). Therefore, the shift in $\theta_s$ is not the main factor driving the change in $H_0$. Instead, in the case of temperature data alone, it provides a handle to shift the peak position that is degenerate with other effects, such as the one of extra relativistic species $\Delta N_{\rm eff}$, which modifies the value of the sound horizon $r_s$.
As a result of the strong interaction, the constraints to $N_{\rm eff}$ and $H_0$ become significantly weakened in an analysis of TT data alone, and the value of the Hubble constant can be made compatible with the S$H_0$ES measurement (see red dashed line). 
The requirement to maintain the matter-radiation equality redshift $z_{\rm eq}$ fixed while increasing $N_{\rm eff}$ results in significant degradation of the constraints on $\omega_{\rm cdm}$.
Moreover, the impact of the strong interaction on the amplitude and tilt of the CMB can be compensated by adjusting $A_s$ and $n_s$, leading to a significant shift in these parameters. This also allows for significantly reducing constraints on neutrino masses. 
Interestingly, the overall impact on the matter power spectrum is to decrease the cosmological parameter $S_8$, thereby opening up the possibility of simultaneously resolving the $S_8$ and $H_0$ tension.

\begin{figure*}
    \centering
    \includegraphics[width=1\linewidth]{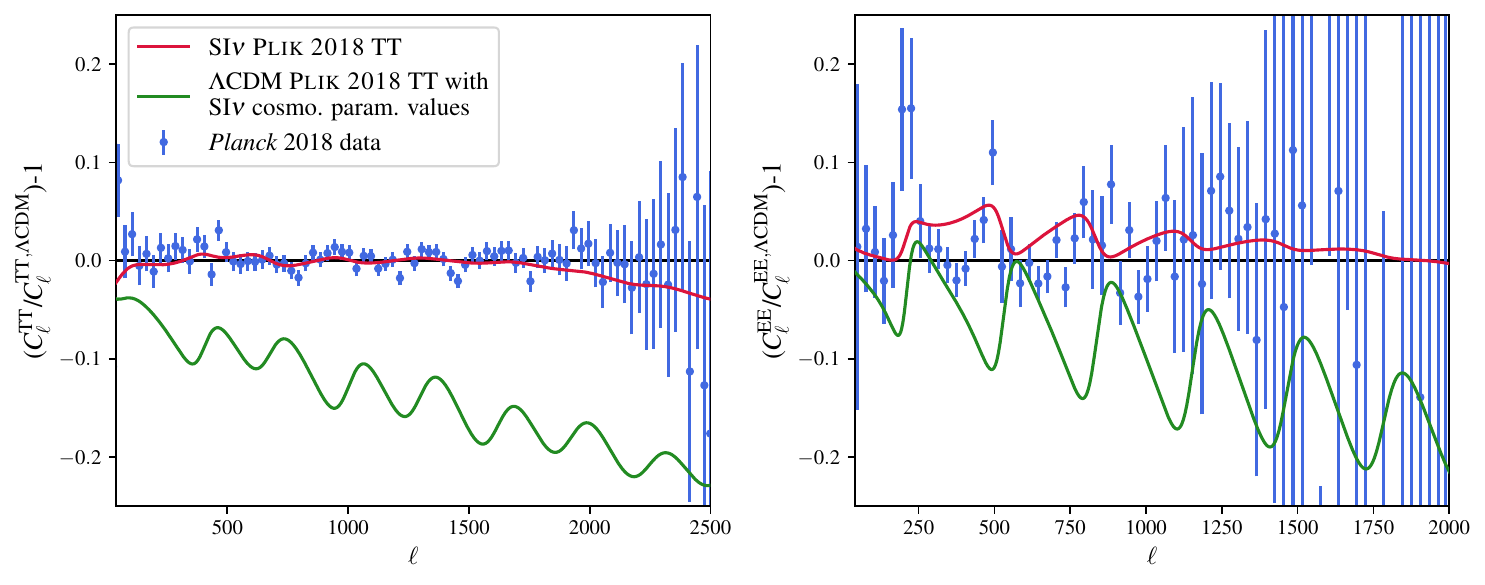}
    \caption{Comparison between the {\sc Plik} 2018 CMB power spectra in the SI$\nu$ mode (red) and in $\Lambda$CDM (green) with values of cosmological parameters from the SI$\nu$ best-fits, together with the {\sc Plik} 2018 data points all normalized to $\Lambda$CDM. We show the temperature power spectrum (TT) on the left and the EE power spectrum on the right.}
    \label{fig:residuals}
\end{figure*}

We illustrate the impact of neutrino strong self-interactions in \cref{fig:residuals}. We depict in green a realization of the $\Lambda$CDM model where all the cosmological parameters have been adjusted to their corresponding best-fit values in the SI$\nu$ model (shown in red for comparison) considering solely TT data and treating neutrinos as free-streaming. All curves are normalized to the $\Lambda$CDM cosmology. From the depiction of the TT data in the left panel, one can see that this model would be firmly excluded (green line) but becomes compatible with TT data when the interactions are included (red line).
On the other hand, the right panel highlights that EE data remains sensitive to the SI$\nu$ mode. This is because EE data is not affected by the same level of degeneracies between the various contributions to the power spectrum (Sachs-Wolfe, integrated Sachs-Wolfe and Doppler terms in the case of TT), and thus shows much sharper acoustic peaks \cite{Zaldarriaga:2003bb,Galli:2014kla}. It is thus much harder to significantly shift the acoustic peaks in the EE power spectrum without being ruled out by {\sc Plik} data.

\begin{figure*}
    \centering
    \includegraphics[width=\linewidth]{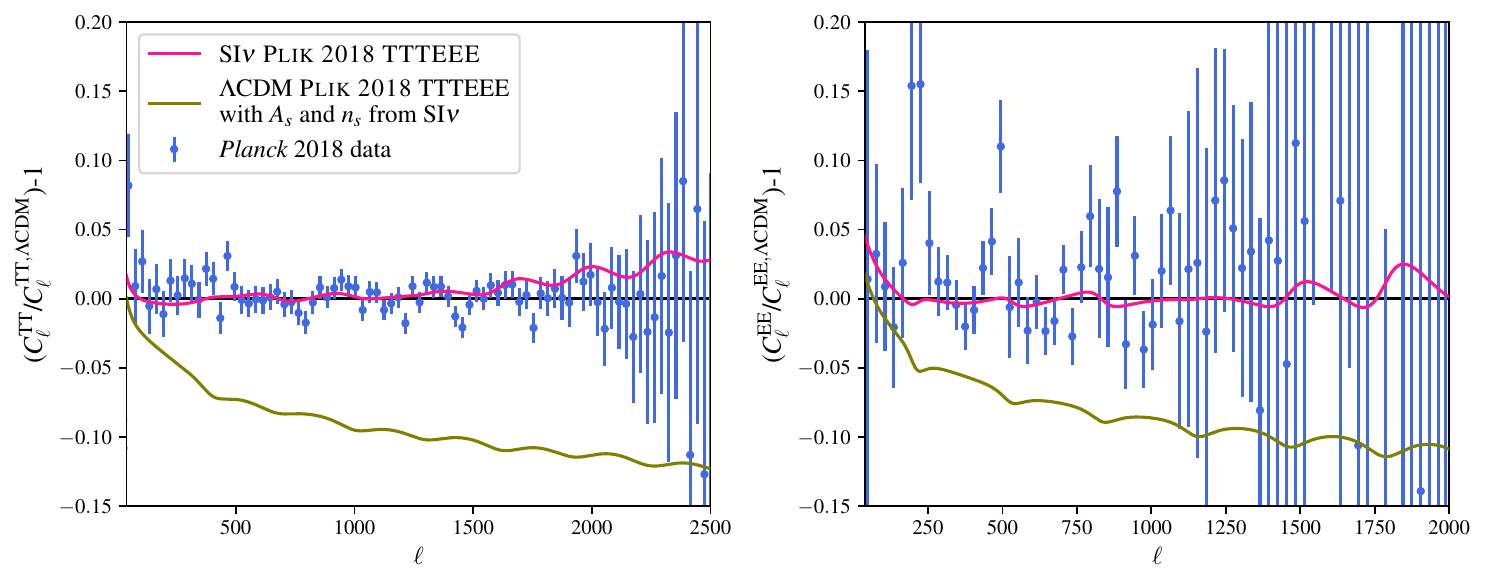}
    \caption{Comparison between the {\sc Plik} 2018 CMB power spectra in the SI$\nu$ mode (pink) and in $\Lambda$CDM (olive green) with values of $A_s$ and $n_s$ from the SI$\nu$ best-fits, together with the {\sc Plik} 2018 data points all normalized to $\Lambda$CDM. We show the temperature power spectrum (TT) on the left and the EE power spectrum on the right.}
    \label{fig:residuals_As_ns}
\end{figure*}

Consequently, when including polarization data in the analysis, the SI$\nu$ mode with large $H_0$ and low $S_8$ is excluded (solid red line in \cref{fig:TT-vs-TTTEEE}). However, large self-interactions remain possible by exploiting the degeneracy with $A_s$ and $n_s$, which remains present even with EE data included. Notably, this is not accompanied by a significant shift in $\Delta N_{\rm eff}$ and $H_0$.

We illustrate this bimodality at the level of the power spectra in \cref{fig:residuals_As_ns}. We show in olive green a realization of the $\Lambda$CDM model where the values of $A_s$ and $n_s$ have been adjusted to their best-fit values in the SI$\nu$ model when analyzing TTTEEE data (shown in pink for comparison) and treating neutrinos as free-streaming. This model would be strongly excluded by both TT and EE data but remains allowed by both datasets when the interactions are switched on.
Therefore, while suppressing the neutrino free-streaming to alter the neutrino phase-shift imprinted in the CMB is not feasible, reducing it sufficiently can significantly impact the reconstructed posterior of $A_s$ and $n_s$.

New polarization data hold great interest in confirming the findings of {\sc Plik} 2018, considering the potential role of neutrino self-interactions in addressing cosmic tensions. In fact, the results from ACT DR4 suggest some level of discrepancy with the {\sc Plik} results \cite{Kreisch:2022zxp}. We discuss updated constraints with NPIPE data in \cref{sec:CMB}.
Furthermore, other cosmological probes sensitive to $A_s$ and $n_s$ may help break the degeneracy between the MI$\nu$ and SI$\nu$ models. This is precisely the role of the LSS data analyzed with the EFTofLSS, which we discuss in \cref{sec:LSS}.

\section{Updated \textit{Planck} and BAO data analysis}
\label{sec:CMB}

In this section, we update the constraints for the self-interacting neutrino model described in \cref{model} in light of updated {\it Planck} and DESI BAO data, using both the Markov Chain Monte Carlo algorithm (MCMC) and profile likelihoods. The comparison of Bayesian and Frequentist methods will allow us to test for prior dependence and provide us with a simple test of model preference, as done in \cite{Camarena_2023,Camarena_2025}.
The comparison with LSS full-shape data and the combined analyses are presented in \cref{sec:LSS}.

\subsection{Datasets}

Our datasets are defined as follows:

\begin{itemize}

\item \textsc{\bf Plik 2018}: The TT, TE, and EE CMB power spectrum data from \textit{Planck} 2018, described through the {\sc Plik} likelihood \cite{Planck:2018vyg}. We also include the PR3 lensing and low-$\ell$ TT and EE likelihoods. 

\item  \textsc{\bf CamSpec NPIPE}: The TT, TE, EE CMB power spectrum data from \textit{Planck} 2020 ({\sc NPIPE})  \cite{1807.06209}, described using the {\sc CamSpec} likelihood\footnote{An alternative likelihood called {\sc Hillipop} is also available \cite{Tristram:2023haj}.} \cite{Rosenberg:2022sdy}. As for \textsc{Plik 2018},  we also include the PR3 lensing and low-$\ell$ TT and EE likelihoods.

\item  \textbf{Pantheon$^+$}: (Uncalibrated) luminosity distance measurements from Type Ia supernovae (SNe Ia) as compiled by the Pantheon+ team \cite{brout_pantheon_2022}.

\item  \textbf{BAO BOSS}: Baryonic Acoustic Oscillations (BAO) measurements of the DR12 galaxy sample of BOSS \cite{BOSS:2016wmc} from SDSS-III, low redshift BAO measurements from the 6dF Galaxy Survey \cite{Beutler:2011hx}, and clustering from the SDSS DR7 Main Galaxy Sample I \cite{ross_clustering_2015}.

\item  \textbf{BAO DESI}: The DESI BAO data presented in Refs.~\cite{DESI:2024uvr,DESI:2024mwx,DESI:2024lzq}. More precisely, we use the data listed in Tab.~1 of Ref.~\cite{DESI:2024mwx}, which spans a redshift range of $z\sim 0.1-4.1$. These data comprise a compilation of low redshift galaxies from the bright galaxy survey, luminous red galaxies, emission line galaxies, quasars, and Lyman$-\alpha$ forest quasars, which trace the distribution of neutral hydrogen. 
\end{itemize}

\subsection{Analysis methods}

We conduct both Bayesian and Frequentist analyses, using \textsc{MontePython-v3} \cite{Brinckmann:2018cvx} interfaced with our modified \texttt{CLASS} code. We run the Metropolis-Hastings algorithm, considering our chains to be converged when the Gelman-Rubin criterion  \cite{Gelman:1992zz} reaches $R-1 < 0.01$ for all the sampled parameters.
All parameters are varied within broad flat priors, except for $G_{\rm eff}$ for which we employ a logarithmic prior. To tackle the bimodality of the posterior distribution of $G_{\rm eff}$, we follow the approach of previous works \cite{Kreisch:2019yzn,Camarena_2023,Camarena_2025} and divide the parameter space of $G_{\rm eff}$ into two distinct regimes: the moderately interacting (MI$\nu$) mode, spanning from $10^{-5}$ to $10^{-2.3}$, and the  strongly-interacting (SI$\nu$) mode spanning from $10^{-2.3}$ to $10^{-0.3}$. We note that the MI$\nu$ mode effectively becomes $\Lambda$CDM for small values of $G_{\rm eff}$, while the SI$\nu$ mode does not asymptote to $\Lambda$CDM under any limit.

For the profile likelihood, we fix the value of $\rm log_{10} G_{\rm eff, \nu}$ while varying all the other cosmological parameters within a wide range. We compute the best-fit for 30 values of the coupling constant using a simulated annealing method similar to that described in Refs.~\cite{Schoneberg:2021qvd,Karwal:2024qpt}. For $\rm log_{10} G_{\rm eff, \nu} \in [-2.3, -0.3]$ (SI$\nu$), we use a step size of $\Delta G_{\rm eff}=0.1$, while for $\rm log_{10} G_{\rm eff, \nu} \in [-5.0, -2.3]$ (MI$\nu$), since the profile is flat, we use a larger step size of $\Delta G_{\rm eff}=0.3$. 

For model comparison purposes, we perform the same set of analyses with the $\Lambda$CDM model, assuming one massive and two massless neutrinos with $m_\nu=0.06$ and a total $N_{\rm eff}=3.044$. To ease the presentation of the results, we will refer to the quantity 
\begin{equation}
    \Delta \chi^2(G_{\rm eff, \nu}) =\chi^2(G_{\rm eff, \nu})-\chi^2_{{\rm min},\Lambda \rm CDM} \mathcomma
\end{equation}
as the `profile likelihood'. In this expression, $\chi^2(G_{\rm eff, \nu})$ and $\chi^2_{{\rm min},\Lambda \rm CDM} $ represent the best-fit $\chi^2$ for the self-interacting neutrino model, with $G_\nu$ kept fixed, and $\Lambda$CDM, respectively. The true `profile likelihood' is related to this quantity by a constant, $\Delta\chi^2_{\rm min}\equiv\chi^2_{{\rm min},\Lambda \rm CDM} - {\rm min}{(\chi^2(G_{\rm eff, \nu}))}$, where ${\rm min}(\chi^2(G_{\rm eff, \nu}))$ denotes the minimum $\chi^2$ over the full range of $G_{\rm eff, \nu}$. 

We perform our analyses in two distinct steps: in \cref{sec:Plik-vs-NPIPE}, we compare the use of {\sc Plik 2018} with {\sc CamSpec NPIPE}, while keeping {\sc BAO BOSS} data. In \cref{sec:DESI}, we update the BAO data to the {\sc DESI} sample and only consider {\sc CamSpec NPIPE} likelihood.

\begin{figure}
    \centering
    \includegraphics[width=\columnwidth]{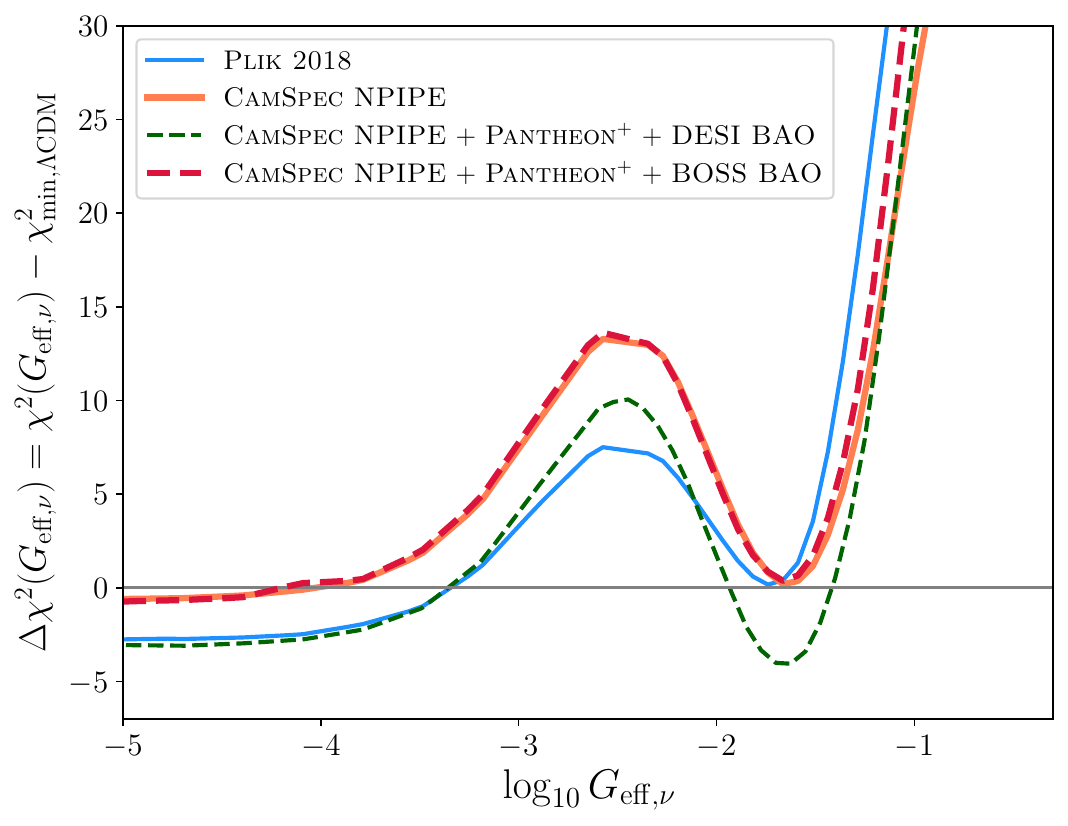}
    \caption{Profile likelihood of $G_{\rm eff}$ from \textsc{Plik 2018} and \textsc{CamSpec NPIPE} with respect to $\Lambda$CDM, with and without the Pantheon$^+$ SN compilation and BAO observations from BOSS or DESI.}
    \label{fig:lcdm profile}
\end{figure}

\renewcommand{\arraystretch}{1.5}
\begin{table*}
 \caption{68\% credible intervals for cosmological parameters (or 95\% for upper limit constraints) obtained when analyzing TT+TE+EE CMB data from \textsc{Plik 2018} and \textsc{CamSpec NPIPE} for the two self-interacting neutrino modes. We also present the $\chi^2$ comparison between the two modes and $\Lambda$CDM, as well as the frequentist confidence interval for $\log_{10}{G_{\rm eff, \nu}/{\rm MeV^{-2}}}$.}
\label{table:CMB parameters}
\begin{tabular}{|c|cc|cc|}
\hline
\multicolumn{5}{|c|}{\sc{\textit{Planck}} TTTEEE}\\
\hline
\hline
&\multicolumn{2}{c}{MI$\nu$} & \multicolumn{2}{|c|}{SI$\nu$}\\
\hline
& \sc{Plik 2018}&\sc{CamSpec NPIPE} & \sc{Plik 2018}  &\sc{CamSpec NPIPE}\\
\hline
$\rm log_{10} G_{\rm eff, \nu} $ & $<$ $-3.49(-4.77)$ &  $<$ $-3.60(-4.92)$ & $-1.7(-1.72)$ $^{+0.17}_{-0.11}$ & $-1.62(-1.64)^{+0.14}_{-0.11}$ \\
$-2\log{\cal L}(\rm log_{10} G_{\rm eff, \nu}) $ & $<$ $-3.2$ & $<-3.31$ & $-1.72^{+0.12}_{-0.16}$ &$-1.64^{+0.12}_{-0.13}$ \\
\hline
$H_0$ & 66.0(66.4)$^{+1.8}_{-1.6}$ & 66.6(67.3) $\pm$ 1.6 & 66.2(66.9)$^{+1.9}_{-1.6}$ & 68.1(68.8)$^{+2.0}_{-2.2}$ \\
$10^{-2}\omega{}_{b }$ & 2.219(2.22)$\pm$ 0.023 & 2.210(2.212)$\pm$ 0.020 & 2.229(2.233)$\pm$ 0.023 &  2.228(2.232)$\pm$ 0.021 \\
$\omega{}_{cdm }$ & 0.1186(0.1171)$\pm$ 0.0030 & 0.1200(0.1189)$\pm$ 0.0031 & 0.1165(0.1515)$^{+0.0028}_{-0.0032}$ & 0.1206(0.1193)$^{+0.0036}_{-0.0047}$\\
$n_{s }$ & 0.9570(0.9577)$\pm$ 0.0084 & 0.9595(0.9602)$\pm$ 0.0078 &  0.9278(0.9295) $\pm$ 0.0076 & 0.9344(0.9348)$\pm$ 0.0080 \\
$10^{9}A_{s }$ & 2.090(2.075)$\pm$ 0.036 & 2.088(2.078)$\pm$ 0.036 & 1.965(1.951)$\pm$ 0.033 & 1.964(1.957)$\pm$ 0.032 \\
$\tau{}_{reio }$ & 0.0543(0.0519)$\pm$ 0.0073 & 0.0532(0.0518)$^{+0.0068}_{-0.0076}$ & 0.0540 (0.0526)$^{+0.0068}_{-0.0076}$ & 0.0537 0.5261)$\pm$ 0.0073 \\
$\sum m_{\nu }$/eVs & $<$ 0.291(0.0006) & $<$ 0.267(0.001) & $<$ 0.289(0.002) &  $<$ 0.289(0.0004) \\
$\Delta N_{\rm{eff}}$ & $-0.13(-0.21)\pm 0.19$ & $-0.01(-0.05)\pm 0.20$ & $-0.30(-0.36)\pm 0.19$ & $-0.03(-0.08)^{+0.22}_{-0.29}$ \\ 
$S_8$
	 & 0.834(0.836)$ ^{+0.012}_{-0.014}$ 
	 & 0.832(0.832)$ ^{+0.010}_{-0.013}$ 
	 & 0.827(0.827)$ \pm 0.013$ 
	 & 0.822(0.826)$ \pm 0.012$ 
	 \\
\hline
 $\Delta \chi^2 =\chi^2_{\rm min}-\chi^2_{{\rm min},\Lambda \rm CDM}$ & $-2.72$ & $-0.6$  & $+0.08$ & $+0.1$\\
 \hline
\end{tabular}
\end{table*}
\renewcommand{\arraystretch}{1.5}

\begin{figure*}
    \centering
    \includegraphics[width=1.73\columnwidth]{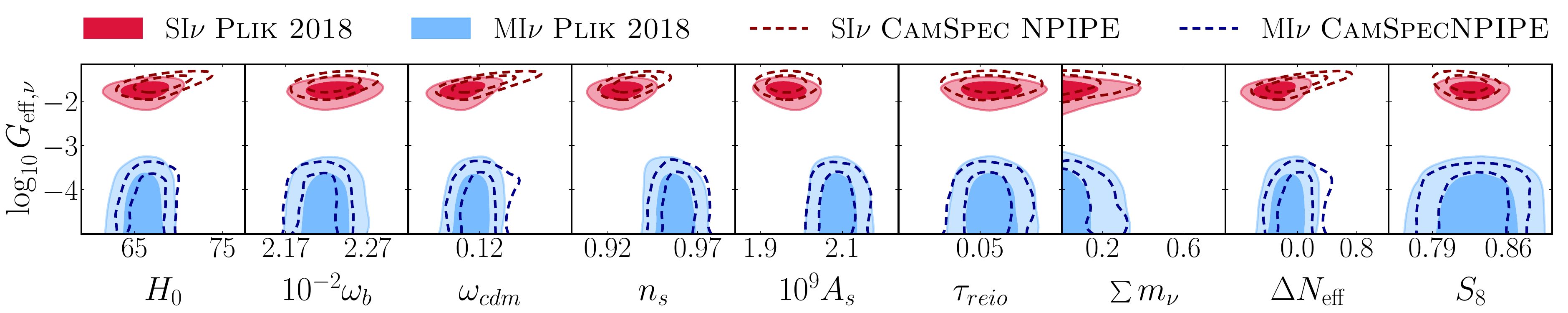}
    \includegraphics[width=0.27\columnwidth]{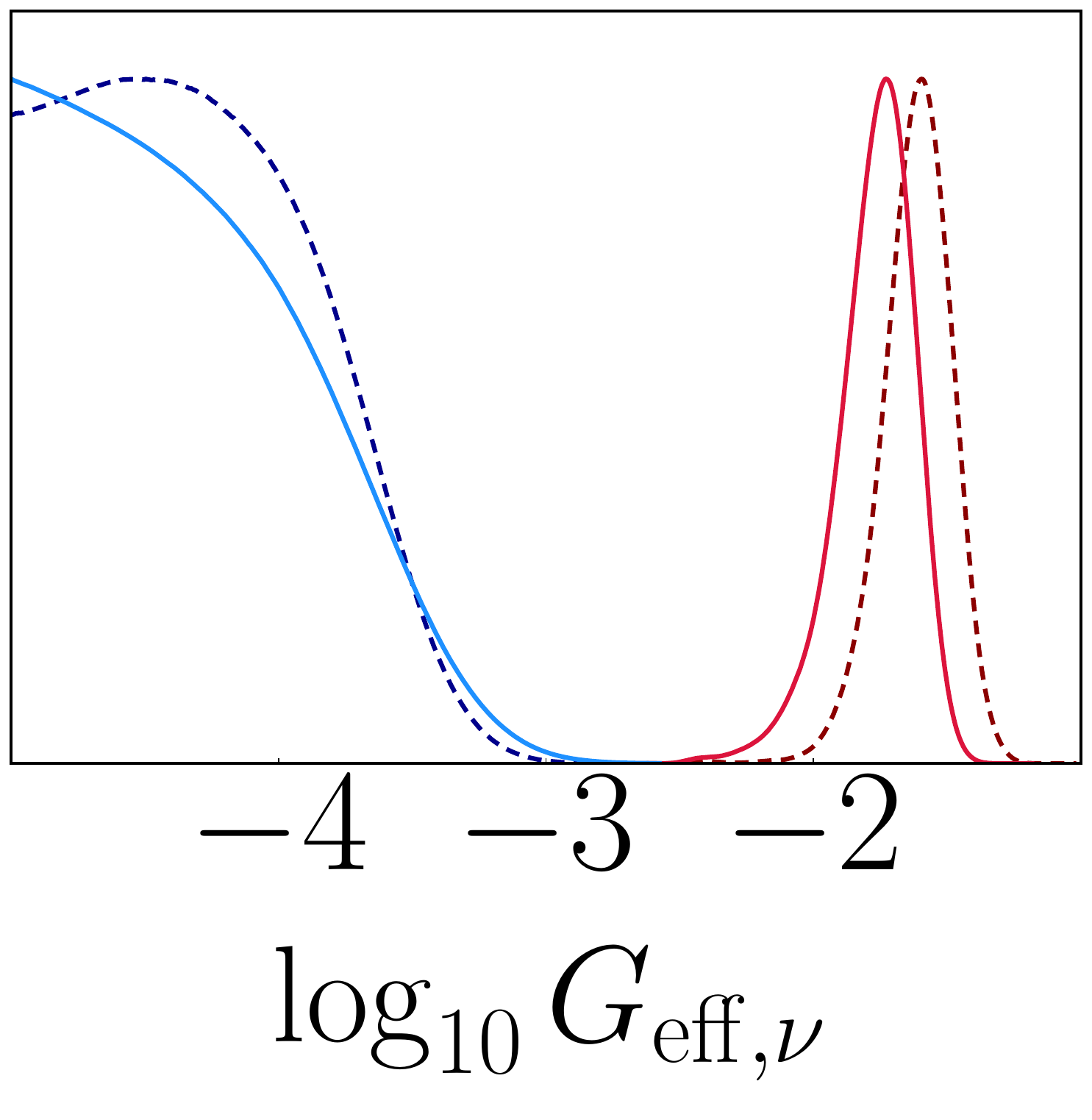}\\
    \caption{1-dimensional and 2-dimensional posteriors of $\log_{10}{G_{\rm eff, \nu}}/{\rm MeV}^{-2}$ for both modes constrained by \textsc{Plik 2018} or \textsc{CamSpec NPIPE}.}
        \label{fig:2-d_NPIPE}
\end{figure*}

\begin{table*}
 \caption{68\% credible intervals for cosmological parameters (or 95\% for upper limit constraints) obtained when analyzing TT+TE+EE CMB data from \textsc{CamSpec NPIPE}, \textsc{Pantheon$^+$}, and \textsc{BAO} data from BOSS and DESI observations,  for the two self-interacting neutrino modes. We also display the $\chi^2$ comparison between both modes and $\Lambda$CDM, as well as the frequentist confidence interval for $\log_{10}{G_{\rm eff, \nu}/{\rm MeV^{-2}}}$.}
\label{table:CMB DESI}
\begin{tabular}{|c|cc|cc|}
\hline
\multicolumn{5}{|c|}{\sc{CamSpec NPIPE+Pantheon$^+$+BAO}}\\
\hline
\hline

&\multicolumn{2}{c}{MI$\nu$} & \multicolumn{2}{|c|}{SI$\nu$}\\
\hline
& \textsc{BOSS BAO}  &\sc{DESI BAO} & \sc{BOSS BAO}  &\sc{DESI BAO}\\
\hline
$\rm log_{10} G_{\rm eff, \nu} $
	 &$ < -3.43(-4.9)$  
	 & $< -3.23(-4.98)$ 
	 & $-1.63(-1.66) ^{+0.13}_{-0.10}$ 
	 & $-1.61(-1.66) ^{+0.14}_{-0.11}$ 
	 \\
$-2\log{\cal L}(\log_{10} G_{\rm eff, \nu})$  & $<-3.36$   & $<-3.25$ & $-1.66 ^{+0.14}_{-0.13}$ & $-1.66 ^{+0.11}_{-0.13}$ \\
\hline
$H_0$
	 & 67.34(67.22)$ ^{+0.88}_{-1.1}$ 
	 & 68.6(68.3)$ ^{+1.1}_{-0.96}$ 
	 & 67.58(67.2)$ ^{+0.91}_{-1.4}$ 
	 & 68.6(68.0)$ ^{+1.2}_{-1.4}$ 
	 \\
$10^{-2}\omega{}_{b }$
	 & 2.216(2.21)$ ^{+0.015}_{-0.017}$ 
	 & 2.229(2.224)$ \pm 0.015$ 
	 & 2.223(2.218)$ ^{+0.015}_{-0.017}$ 
	 & 2.232(2.222)$ ^{+0.015}_{-0.017}$ 
	 \\
$\omega{}_{cdm }$
	 & 0.1198(0.1187)$ ^{+0.0025}_{-0.0031}$ 
	 & 0.1207(0.1197)$ \pm 0.0028$ 
	 & 0.1201(0.1184)$ ^{+0.0030}_{-0.0042}$ 
	 & 0.1209(0.1182)$ ^{+0.0036}_{-0.0044}$ 
	 \\
$n_{s }$
	 & 0.9613(0.9594)$ ^{+0.0061}_{-0.0068}$ 
	 & 0.9668(0.965)$ \pm 0.0063$ 
	 & 0.9327(0.9299)$ ^{+0.0054}_{-0.0064}$ 
	 & 0.9360(0.9317)$ ^{+0.0057}_{-0.0064}$ 
	 \\
$10^9 A_s$
	 & 2.089(2.076)$ \pm 0.033$ 
	 & 2.104(2.09)$ \pm 0.032$ 
	 & 1.960(1.953)$ ^{+0.029}_{-0.032}$ 
	 & 1.963(1.949)$ ^{+0.029}_{-0.032}$ 
	 \\
$\tau{}_{reio }$
	 & 0.0538(0.0518)$ ^{+0.0067}_{-0.0075}$ 
	 & 0.0562(0.0534)$ ^{+0.0062}_{-0.0072}$ 
	 & 0.0529(0.0518)$ \pm 0.0073$ 
	 & 0.0531(0.0512)$ \pm 0.0072$ 
	 \\
$\sum m_{\nu}$/eVs
	 & $< 0.136(0.002)$
	 &  $< 0.0934(0.0002)$
	 & $< 0.221(0.061)$ 
	 & $< 0.179(0.005)$
	 \\
$\Delta N_{\rm{eff}}$
	 & $0.01(-0.08) ^{+0.14}_{-0.20}$ 
	 & 0.12(0.04)$ ^{+0.18}_{-0.16}$ 
	 & $-0.08(-0.18) ^{+0.16}_{-0.25}$ 
	 & 0.01(-0.17)$ ^{+0.20}_{-0.26}$ 
	 \\
$S_8$
	 & 0.830(0.832)$ \pm 0.011$ 
	 & 0.827(0.825)$ ^{+0.010}_{-0.012}$ 
	 & 0.824(0.829)$ \pm 0.012$ 
	 & 0.822(0.828)$ \pm 0.011$ 
	 \\
\hline
$\Delta \chi^2 =\chi^2_{\rm min}-\chi^2_{{\rm min},\Lambda \rm CDM}$  & $-0.72$& $-3.1$ &  $+0.36$
 &$-4.3$\\
 \hline
\end{tabular}
\end{table*}

\begin{figure*}
    \includegraphics[width=1.7\columnwidth]{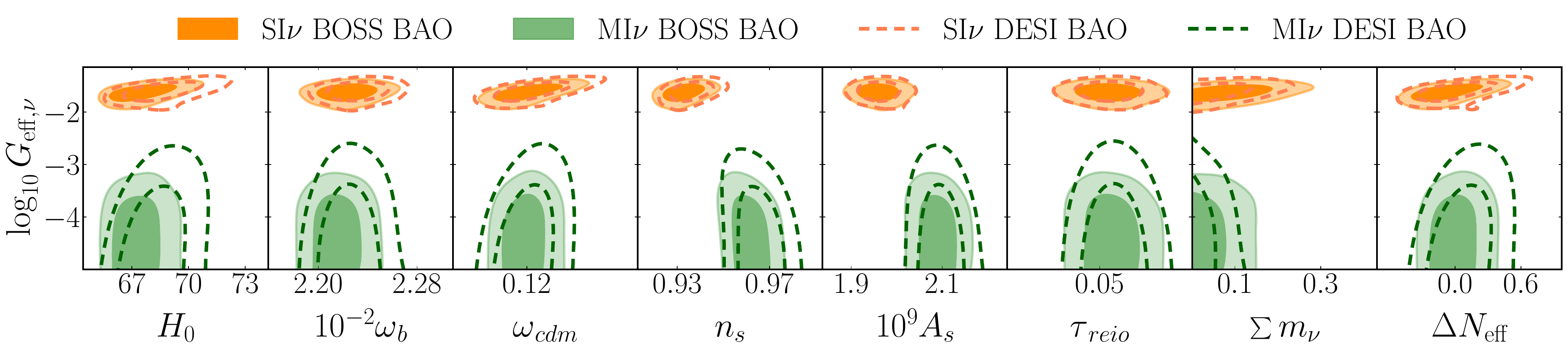}
     \includegraphics[width=0.3\columnwidth]{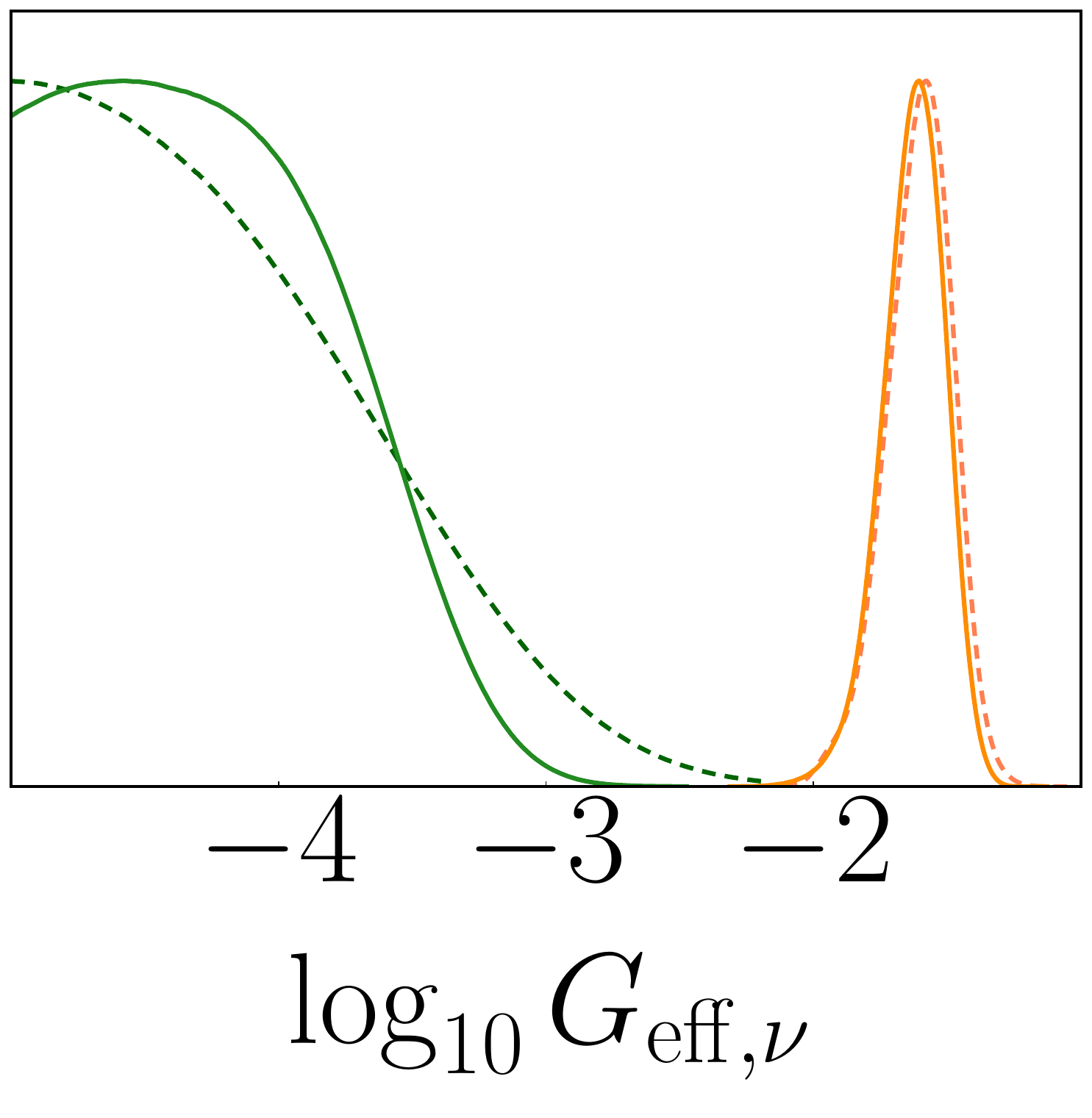}\\
    \caption{1-dimensional and 2-dimensional posteriors of $\log_{10}{G_{\rm eff, \nu}}/{\rm MeV}^{-2}$ for both modes constrained by \textsc{CamSpec NPIPE} and \textsc{Pantheon$^+$}, as well as BAO observations from BOSS or DESI.}
    \label{fig:2-d_DESI}
\end{figure*}

\subsection{CMB constraints}
\label{sec:Plik-vs-NPIPE}

The reconstructed parameter posteriors for the corresponding modes and datasets are provided in \cref{table:CMB parameters}. We also report the Frequentist confidence interval for $\log_{10}G_{\rm eff}$, as well as the $\Delta\chi^2$ for the global minimum within each mode. 
In \cref{fig:2-d_NPIPE}, we compare the 1D and 2D posteriors of $\log_{10}{G_{\rm eff, \nu}}/{\rm MeV}^{-2}$ reconstructed from \textsc{Plik 2018} or \textsc{CamSpec NPIPE}. We also show the $\Delta\chi^2$  with respect to $\Lambda$CDM as a function of $\rm log_{10} G_{\rm eff, \nu}$ for each dataset in \cref{fig:lcdm profile}. 

First, one can see that the SI$\nu$ mode persists in the new \textsc{CamSpec NPIPE}, with only a mild shift towards higher values of $\log_{10}{G_{\rm eff, \nu}/{\rm MeV^{-2}}}= -1.62^{+0.14}_{-0.11}$ compared to the \textsc{Plik 2018} analysis. However, the constraints on the coupling constant slightly increase in the MI$\nu$ mode, with  $\log_{10}{G_{\rm eff, \nu}/{\rm MeV^{-2}}}< -3.60$ (at 95\% C.L.), compared to $\log_{10}{G_{\rm eff, \nu}/{\rm MeV^{-2}}}< -3.49$ for \textsc{Plik 2018}.

These results are corroborated by the `profile likelihood' (normalized to $\Lambda$CDM) reported in \cref{fig:lcdm profile}. One can discern the presence of the two modes and use this figure to assess whether one mode is favored over the other. Surprisingly, while we confirm that the \textsc{Plik 2018} data favor the mildly interacting mode compared to the strongly-interacting mode \cite{Kreisch:2019yzn,Camarena_2023,Camarena_2025}, with $\Delta\chi^2({\rm MI\nu}-{\rm SI\nu})\equiv\chi^2({\rm MI\nu})-\chi^2({\rm SI}\nu)=-2.8$, the \textsc{CamSpec NPIPE} data provide less decisive preferrence, with $\Delta\chi^2({\rm MI\nu}-{\rm SI\nu})=-0.7$.  
This finding is surprising considering the higher statistical power of the \textsc{CamSpec NPIPE} likelihood compared to \textsc{Plik 2018}. 
While one may have anticipated that using the latest CMB data would result in the SI$\nu$ mode being evidently disfavored, here we find that the possibility of strongly-interacting neutrinos remains open. However, it is important to emphasize that neither mode is statistically significantly preferred over $\Lambda$CDM.

To gauge further the impact of \textsc{CamSpec NPIPE}, we present in \cref{fig:2-d_NPIPE} the two-dimensional posteriors of the coupling constant against other cosmological parameters for each dataset in both modes. 
This confirms that \textsc{NPIPE} data is not more constraining than the \textsc{Plik 2018} data for this model, allowing larger values of $\Delta N_{\rm eff}$, and larger values (at about $1\sigma$) for the parameters it correlates with, namely $H_0$, $\omega_{\rm cdm}$ and $n_s$. Similar results were found for the cases of free-streaming or pure fluid, as well as extra radiation (on top of regular neutrinos) in Refs.~\cite{Allali:2024cji,Saravanan:2025cyi}. 

 Similarly to \textsc{Plik 2018}, the bimodality in $\log_{10}{G_{\rm eff, \nu}}/{\rm MeV}^{-2}$ obtained with {\sc CamSpec NPIPE} is correlated with $A_s$ and $n_s$, which also exhibit a bimodality with a $1-2\sigma$ difference between each mode.
However, despite the shift in $\log_{10}{G_{\rm eff, \nu}}/{\rm MeV}^{-2}$ to higher values, there is no bimodality in $H_0$ and $S_8$. We confirm with updated data that the self-interacting neutrino model studied here cannot resolve the Hubble or $S_8$ tensions.

\subsection{BAO constraints and impact on $\sum m_\nu$}
\label{sec:DESI}
We now incorporate BAO and (uncalibrated) SN1a data into the analysis, aiming to quantify the impact of the updated BAO data on the model. DESI data have been found to favor slightly different values of the product $H_0r_s$ compared with SDSS \cite{DESI:2024mwx}, which could potentially impact constraints on self-interacting neutrinos, similar to what occurs when considering free-streaming, or pure fluid, extra radiation \cite{Allali:2024cji,Saravanan:2025cyi}. We present the results of analyses that include either BOSS BAO+Pantheon$^+$ or DESI BAO+Pantheon$^+$ along with {\sc CamSpec NPIPE}, in \cref{table:CMB DESI} (where we provide constraints for all cosmological parameters). We also present the 1D and 2D posteriors of $\log_{10}{G_{\rm eff, \nu}}/{\rm MeV}^{-2}$ against the parameters of interest in \cref{fig:2-d_DESI}, and the profile likelihoods in \cref{fig:lcdm profile}.

We find that BOSS BAO+Pantheon$^+$ data have a minimal impact on the overall results, with only a small increase in the preference for the MI$\nu$ mode over the SI$\nu$ mode, although they remain statistically compatible with $\Delta\chi^2({\rm MI\nu}-{\rm SI\nu})=-1.08$. In addition, there is still no preference over $\Lambda$CDM (as depicted in \cref{fig:lcdm profile}). In contrast, the DESI BAO data significantly influences the results and, in fact, leads to the  SI$\nu$ mode being (slightly) favored with $\Delta\chi^2({\rm MI\nu}-{\rm SI\nu})=+1.2$. Both MI$\nu$ and SI$\nu$ modes appear to fit the data slightly better than $\Lambda$CDM with $\Delta\chi^2$(MI$\nu$)$=-3.1$ and $\Delta\chi^2$(SI$\nu$)$=-4.3$, respectively (as shown in \cref{fig:lcdm profile}). However, these small improvements are inconclusive due to the inclusion of three additional parameters in our analysis. This aligns with the recent findings of the DESI collaboration \cite{Whitford:2024ecj}, which finds a potential hint for non-standard neutrino properties when measuring the phase-shift induced by free-streaming species on the BAO peak. Although we cannot directly compare our analyses, it confirms that different neutrino properties (or some additional exotic relativistic species) appear to be slightly favored by DESI data.

Furthermore, it is worth noting that the constraints on the neutrino mass are relaxed in the self-interacting neutrino scenario compared to the non-interacting scenario. This is interesting given the very strong bound found by the DESI collaboration, $\sum m_\nu <0.072$ eV in the flat $\Lambda$CDM model. Here, we find $\sum m_\nu< 0.0934$ eV for the MI$\nu$ mode and $\sum m_\nu< 0.179$ eV for the SI$\nu$ mode. 
Thus, it will be interesting to revisit the analysis of self-interacting neutrinos with future DESI data, as this model may provide a potential avenue for reconciling laboratory constraints with cosmological bounds. 
\section{updated large-scale structure constraints}
\label{sec:LSS}

\subsection{Large-Scale Structures: theory and data} 

We now turn to performing the same analysis for BOSS and eBOSS full-shape data analyzed with the EFT and Big Bang Nucleosynthesis (BBN) data. We make use of the following datasets: 

 \begin{itemize}
 \item  \textbf{EFTofBOSS}: The galaxy power spectra (monopole and quadrupole) from BOSS DR12 luminous red galaxies (LRG), cross-correlated with the reconstructed BAO parameters \cite{BOSS:2015fqm}. The SDSS-III BOSS DR12 galaxy sample data and covariances are described in Refs.~\cite{BOSS:2016wmc,Kitaura:2015uqa}. The measurements, obtained in Ref.~\cite{Zhang:2021yna}, are from BOSS catalogs DR12 (v5)~\cite{BOSS:2015ewx}. They are divided into four skycuts, made up of two redshift bins, namely LOWZ with $0.2<z<0.43 \  (z_{\rm eff}=0.32)$, and CMASS with $0.43<z<0.7  \ (z_{\rm eff}=0.57)$, with north and south galactic skies for each denoted NGC and SGC, respectively.

\item  \textbf{EFTofeBOSS}: The galaxy power spectra (monopole and quadrupole) from eBOSS DR16 quasi-stellar objects (QSO) \cite{eBOSS:2020yzd}. The covariances of the galaxy power spectra are built from the EZ-mocks described in Ref.~\cite{Chuang:2014vfa}, while the QSO catalogs are detailed in Ref.~\cite{eBOSS:2020mzp}. These data comprise 343 708 quasars selected in the redshift range $0.8<z<2.2$ ($z_{\rm eff}=1.52$), divided into two skies, NGC and SGC~\cite{Beutler:2021eqq,eBOSS:2020gbb}. The EFTofeBOSS likelihood is described in Ref.~\cite{Simon:2022csv}.

\item  \textbf{BBN}: BBN measurement of $\omega_b$ \cite{Schoneberg:2019wmt} that incorporates the theoretical prediction of Ref.~\cite{Consiglio:2017pot}, along with the experimental fraction of deuterium and helium from Refs.~\cite{Cooke:2017cwo} and~\cite{Aver:2015iza}, respectively.

\item \textbf{Ly-$\alpha$ prior}: Gaussian priors on the amplitude $\Delta_L^2$ and slope $n_L$ of the matter power spectrum at a redshift $z = 3$ and wavenumber $k = 0.009$ s/km $\sim 1~h$/Mpc, as defined in Refs.~\cite{Goldstein:2023gnw,Rogers:2023upm} and extracted from the eBOSS 1D Ly-$\alpha$ power spectrum \cite{eBOSS:2018qyj}.\\

 \end{itemize}
BOSS and eBOSS data have been extensively used in the literature to study neutrino properties, such as their mass, number of species and non-standard interactions (see \textit{e.g.}, \cite{Chudaykin:2019ock,Ivanov:2019hqk,Colas:2019ret,Kumar:2022vee,Simon:2022csv,Racco:2024lbu}).
In addition, a preference for the SI$\nu$ mode in the EFTofBOSS data has been highlighted in recent literature, namely in Refs.~\cite{He:2023oke,Camarena_2023,Camarena_2025}. While it is particularly interesting that both CMB and LSS data would allow (and potentially favor) the SI$\nu$ mode, the location of the SI$\nu$ posterior distribution within EFTofBOSS seems to differ from that observed in the CMB data \cite{Camarena_2025}.
 
In this study, we make use of BOSS data\footnote{Although recent investigations of DESI data analyzed with the EFTofLSS have appeared \cite{DESI:2024hhd}, DESI data are not publicly available to perform the same analysis.} analyzed with the \texttt{PyBird} code \cite{DAmico:2020kxu}, an alternative to the \texttt{CLASS-PT} code \cite{Chudaykin:2020aoj} employed in previous studies of interacting neutrino models to perform the EFTofBOSS analyses. While both codes have demonstrated good agreement when identical analysis choices are made \cite{Simon:2022lde,Holm:2023laa}, it is interesting to assess the robustness of the results with different implementations and choices for the EFT analysis.
In addition, we will include the EFTofeBOSS data for the first time in this context, a dataset known to help constrain neutrino properties by providing additional cosmological information at intermediate redshifts.\footnote{Ref.~\cite{Simon:2022csv} shows how an EFT analysis of the eBOSS data improves the constraints on the sum of the neutrino mass by $40 \%$  and the constraints on the effective number of relativistic species by $30 \%$ compared to BOSS (+BBN) alone.} We will then compare our LSS full-shape results with the \textsc{CamSpec NPIPE} analysis (which yields slightly different results than the \textsc{Plik 2018} results as we have shown in \cref{sec:CMB}) to assess whether a consistent picture of neutrino interaction emerges between updated CMB and LSS data.

\subsection{EFT likelihood}

In the following, we provide all the crucial details about the \texttt{PyBird} EFT likelihood we applied to the (e)BOSS data:

\begin{itemize}
    \item \textbf{Priors:} In this paper, we use the ``West Coast'' (WC) priors for the EFT parameters~\cite{DAmico:2019fhj,Colas:2019ret,DAmico:2020kxu}, unlike previous analyses~\cite{He:2023oke,Camarena_2023,Camarena_2025} that used the ``East Coast'' (EC) priors~\cite{Philcox:2021kcw} (see Refs.~\cite{Simon:2022lde,Holm:2023laa} for details on the EFT priors). In total, we vary seven EFT parameters per skycut: three bias parameters ($b_1$, $b_3$, and $c_2 \equiv (b_2+b_4)/\sqrt{2}$), two counterterm parameters ($c_{\rm ct}$ and $c_{r,1}$), and two stochastic parameters ($c_{\epsilon, 0}$ and $c_{\epsilon}^{\rm quad}$).
    It is worth noting that the BOSS and eBOSS data have 4  and 2 skycuts, respectively, which means we vary a total of 42 EFT parameters in our analysis. Following Ref.~\cite{DAmico:2020kxu}, the EFT parameters that enter linearly in the theory (namely, $b_3$, $c_{\rm ct}$, $c_{r,1}$, $c_{\epsilon, 0}$, and $c_{\epsilon}^{\rm quad}$) are analytically marginalized with a Gaussian prior of width 2 centered on 0. On the other hand, for the remaining parameters, we adopt the method of Ref.~\cite{DAmico:2022osl}: we vary the linear bias parameter $b_1$ within a flat prior $b_1 \in [0,4]$, while we vary $c_2$ within a Gaussian prior of width 2 centered on 0.
    \item \textbf{Renormalization scales:} For the BOSS data, we set $k_\textsc{m} = k_\textsc{nl} = 0.7 \, h {\rm Mpc}^{-1} $, which serves as the renormalization scale controlling the spatial derivative expansion (determined by the typical extension of the host halo), and we set $k_R = 0.35 \, h {\rm Mpc}^{-1}$, which corresponds to the renormalization scale of the velocity products appearing in the redshift-space expansion. For the eBOSS data, we set $k_\textsc{m} = k_\textsc{nl} = 0.7 \, h {\rm Mpc}^{-1} $ and $k_R =  0.25 \, h {\rm Mpc}^{-1}$ (as detailed in Ref.~\cite{Simon:2022csv}).
    In addition, we set the mean galaxy number density to $\Bar{n}_g = 4 \cdot 10^{-4} \, ({\rm Mpc}/h)^3$, based on estimates from Refs.~\cite{DAmico:2019fhj,DAmico:2021ymi} for the BOSS data. For the eBOSS data we use $\Bar{n}_g = 2 \cdot 10^{-5} \, ({\rm Mpc}/h)^3$.
    \item \textbf{Cutoff scales:} For the BOSS data, following Refs.~\cite{Colas:2019ret,DAmico:2020kxu}, we consider $ k \in [0.01, 0.20] \, h {\rm Mpc}^{-1}$ for the LOWZ sample and $k \in [0.01, 0.23] \, h {\rm Mpc}^{-1}$ for the CMASS sample. For the eBOSS data, following Ref.~\cite{Simon:2022csv}, we consider $ k \in [0.01, 0.24] \, h {\rm Mpc}^{-1}$.
    \item \textbf{Observational effects:} Finally, our analysis also incorporates several observational effects~\cite{DAmico:2019fhj}, including the Alcock-Paszynski effect~\cite{Alcock:1979mp}, the window functions as implemented in Ref.~\cite{Beutler:2018vpe} (see App.~A of Ref.~\cite{Simon:2022adh} for more details), and binning~\cite{DAmico:2022osl}.
\end{itemize}
 
 \subsection{LSS constraints}

\begin{figure*}
    \centering
   \includegraphics[width=1.7\columnwidth]{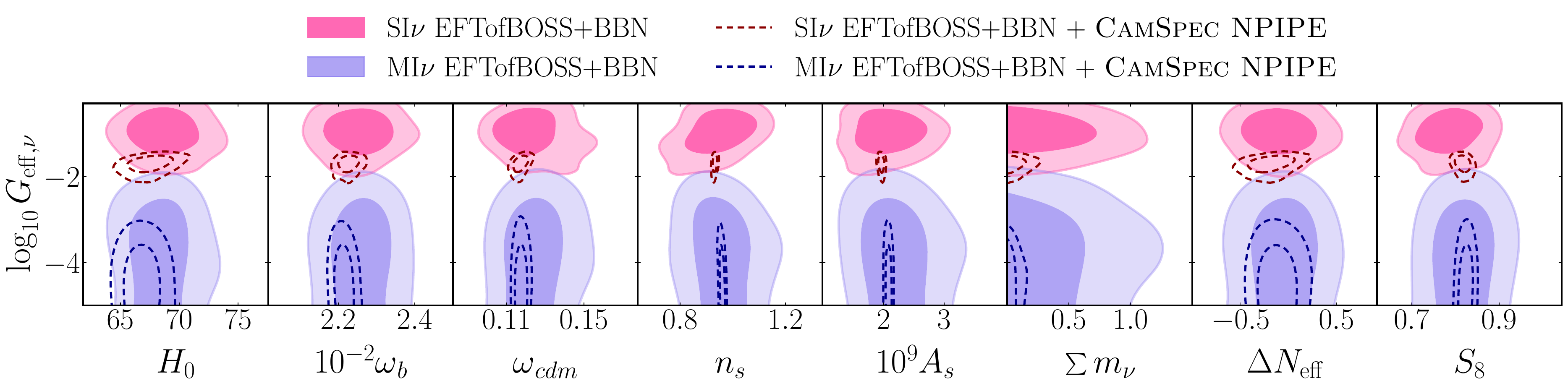}
    \includegraphics[width=0.3\columnwidth]{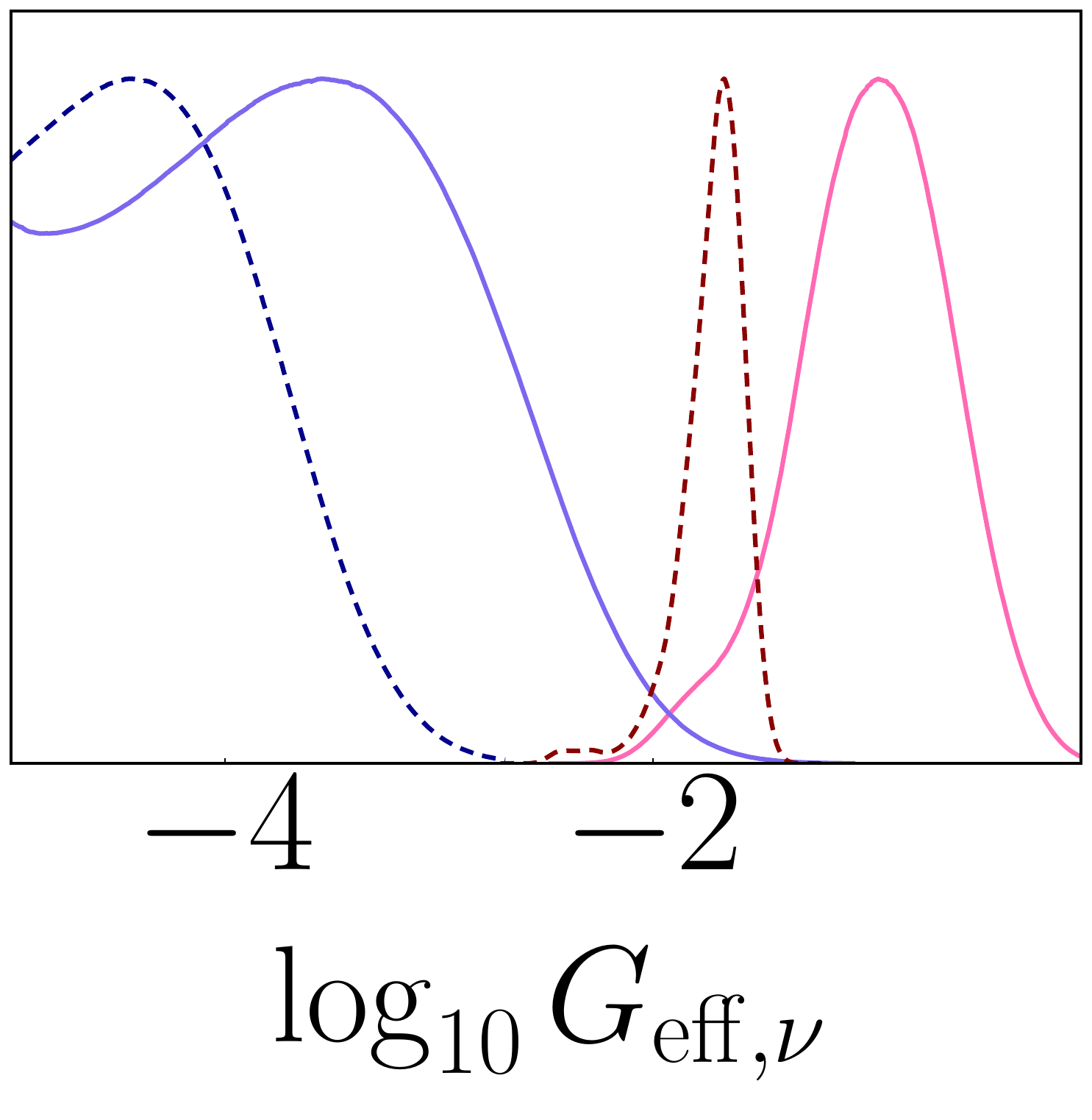}\\
    \caption{1-dimensional and 2-dimensional posteriors of $\log_{10}{G_{\rm eff, \nu}}$ for both modes constrained by EFTofBOSS data with and without \textsc{CamSpec NPIPE}.}
        \label{fig:2-d_EFT}
\end{figure*}

\renewcommand{\arraystretch}{1.5}
\begin{table*}
 \caption{68\% credible intervals for cosmological parameters (or 95\% for upper limit constraints) obtained when analyzing EFTofBOSS+BBN or EFTofBOSS+EFTofeBOSS+BBN datasets for the two self-interacting neutrino modes. We also display the $\Delta\chi^2$ with respect to $\Lambda$CDM, as well as the frequentist confidence interval for $\log_{10}{G_{\rm eff, \nu}/{\rm MeV^{-2}}}$.}
\label{table:BOSS+eBOSS parameters}

\begin{tabular}{|c|cc|cc|}
\hline
\multicolumn{5}{|c|}{\sc{EFTofLSS}}\\
\hline
\hline
&\multicolumn{2}{c}{MI$\nu$} & \multicolumn{2}{|c|}{SI$\nu$}\\
\hline
& EFTofBOSS  &EFTofBOSS+eBOSS & EFTofBOSS  &EFTofBOSS+eBOSS\\
\hline
$\rm log_{10} G_{\rm eff, \nu} $
	 & $<-2.40(-3.13)$
	 & $<-2.47(-3.3) $
	 &$ -0.98(-1.06) ^{+0.39}_{-0.32}$ 
	 & $-0.91(-0.62)\pm 0.33$ 
	 \\
 $-2\log{\cal L}(\log_{10} G_{\rm eff, \nu})$ & unconstrained & unconstrained & $-1.73^{0.45}_{-0.31}$ &$-1.66^{+0.55}_{-0.27}$ \\  
\hline
$H_0$
	 & 68.4(67.7)$ ^{+1.8}_{-2.1}$ 
	 & 68.0(67.2)$ ^{+1.7}_{-1.9}$ 
	 & 68.7(67.8)$ ^{+1.7}_{-2.2}$ 
	 & 68.0(67.2)$ \pm 1.9$ 
	 \\

$10^{-2}\omega{}_{b }$
	 & 2.260(2.245)$ \pm 0.060$ 
	 & 2.268(2.245)$ \pm 0.061$ 
	 & 2.254(2.247)$ \pm 0.059$ 
	 & 2.254(2.238)$ \pm 0.061$ 
	 \\

$\omega{}_{cdm }$
	 & 0.1248(0.1178)$ ^{+0.0095}_{-0.013}$ 
	 & 0.1157(0.11)$ ^{+0.0077}_{-0.010}$ 
	 & 0.1218(0.1127)$ ^{+0.0085}_{-0.014}$ 
	 & 0.1123(0.1066)$ ^{+0.0077}_{-0.0098}$ 
	 \\

$n_{s }$
	 & 0.957(0.935)$ ^{+0.078}_{-0.087}$ 
	 & 1.026(0.991)$ ^{+0.069}_{-0.089}$ 
	 & 0.946(0.959)$ ^{+0.090}_{-0.080}$ 
	 & 1.017(1.065)$ ^{+0.078}_{-0.088}$ 
	 \\

$10^{9}A_{s }$
	 & 2.20(2.13)$ ^{+0.35}_{-0.46}$ 
	 & 2.57(2.52)$ ^{+0.37}_{-0.55}$ 
	 & 2.08(2.4)$ ^{+0.35}_{-0.42}$ 
	 & 2.52(2.98)$ ^{+0.34}_{-0.60}$ 
	 \\

$\sum m_\nu$/eVs
	 & $<$ 1.06(0.0003) 
	 & $<$ 1.11(0.0006)
	 & $<$ 0.926(0.26) 
	 & $<$ 0.991(0.3)
	 \\

$\Delta N_{\rm{eff}}$
	 & $-0.05(-0.12) ^{+0.28}_{-0.25}$ 
	 & $-0.02(-0.13) \pm 0.27$ 
	 & $-0.098(-0.119) \pm 0.25$ 
	 & $-0.10(-0.16) \pm 0.27$ 
	 \\

$S_8$
	 & 0.819(0.838)$ \pm 0.052$ 
	 & 0.843(0.868)$ ^{+0.047}_{-0.055}$ 
	 & 0.787(0.812)$ ^{+0.046}_{-0.052}$ 
	 & 0.813(0.878)$ \pm 0.051$ 
	 \\
\hline
 $\Delta \chi^2 =\chi^2_{\rm min}-\chi^2_{{\rm min},\Lambda \rm CDM}$ & $-0.59$ & $-0.46$  &$-0.26$ & $-1.15$\\
 \hline
\end{tabular}
\end{table*}

\begin{figure*}
    \centering
    \includegraphics[width=\columnwidth]{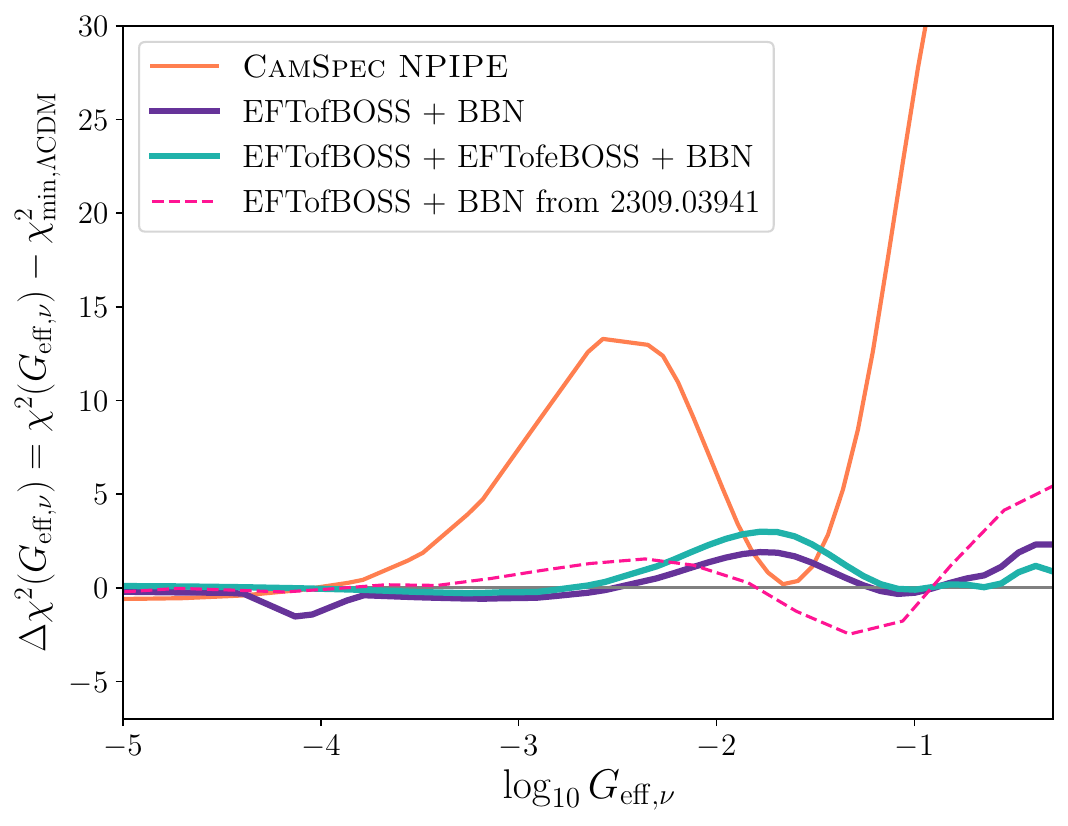}    \includegraphics[width=\columnwidth]{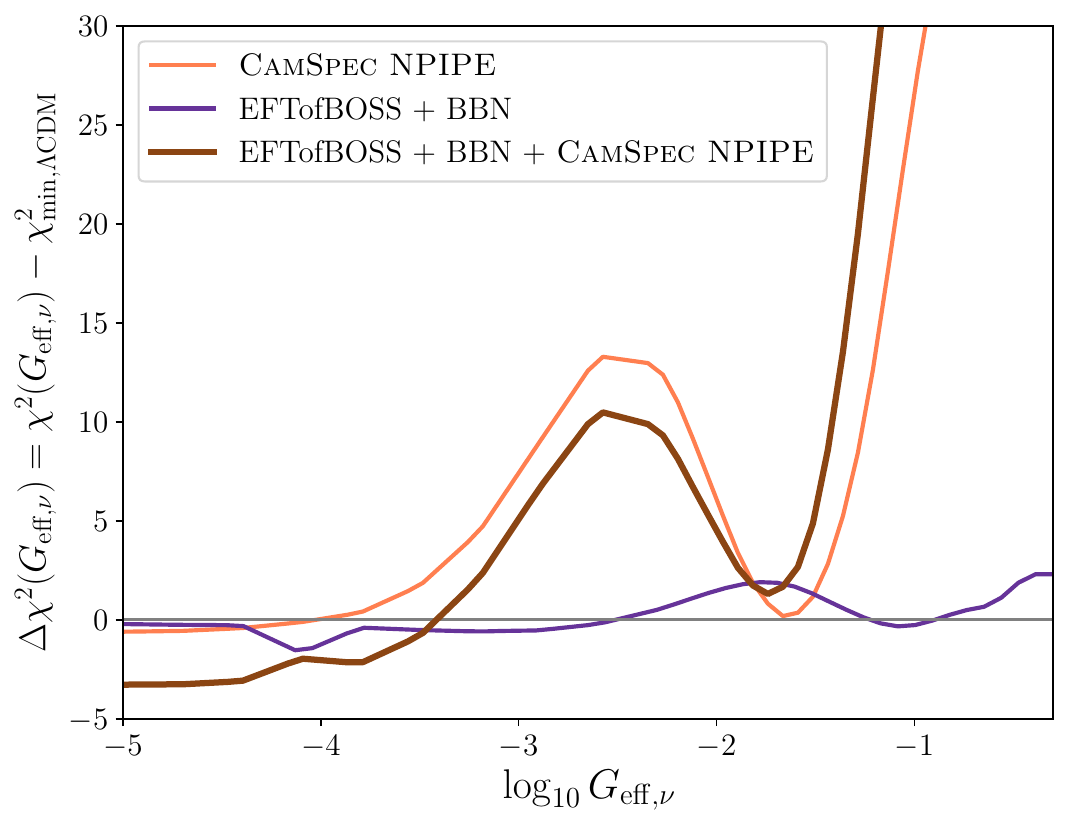}\\
     \caption{{\it Left panel$-$} Profile likelihood (normalized to $\Lambda$CDM) of $G_{\rm eff,\nu}$ for EFTofBOSS + BBN and EFTofBOSS + EFTofeBOSS + BBN compared with that from \textsc{CamSpec NPIPE} and with the full shape analysis from \cite{Camarena_2023}. {\it Right panel$-$} Same as the left panel, now showing the combination of EFTofBOSS and \sc{CamSpec NPIPE}.}
    \label{fig:profile_EFT}
\end{figure*}

\indent Our results are presented in \cref{table:BOSS+eBOSS parameters}, and we show the posteriors for the two interaction modes in \cref{fig:2-d_EFT}, for EFTofBOSS + BBN (full contours). Although the constraints from EFTofBOSS + BBN are weaker than those from \textsc{CamSpec NPIPE}, they also exhibit a (mild) bimodality in \logGeff, in agreement with previous literature \cite{He:2023oke,Camarena_2023,Camarena_2025}.
We further confirm the existence of a small offset in the value of \logGeff for the SI$\nu$ mode between the EFTofBOSS + BBN and {\it Planck} analyses. The use of {\sc CamSpec NPIPE} and a different EFT implementation does not affect these conclusions. \\

In \cref{fig:profile_EFT} (left panel), we present the `profile likelihood' for EFTofBOSS + BBN data\footnote{Here, we follow Ref.~\cite{Holm:2023laa} and promote the EFT priors to effective likelihoods in order to ensure that the EFT parameters remain within physically motivated ranges.} (normalized to the $\Lambda$CDM best-fit $\chi^2$), along the \textsc{CamSpec NPIPE} profile likelihood. Additionally, we report the profile likelihood derived from a different analysis of EFTofBOSS data~\cite{Camarena_2023}. The profile likelihood confirms the offset in the position of the SI$\nu$ mode between CMB data and LSS data. The latter profile exhibits a slight peak precisely where the CMB likelihood reaches a minimum. Furthermore, our EFTofBOSS + BBN profile does not exhibit a preference for either interaction mode. 
This differs from the profiles reported in Refs.~\cite{Camarena_2023,He:2023oke}, which show a small preference for the SI$\nu$ mode. 
In addition, there are significant differences in the values of $A_s$ and $n_s$ between the two analyses, with the EC analyses favoring lower values of $A_s$ and $n_s$ (see \cref{table:BOSSWC_BOSSEC parameters} of \cref{app:EC-vs-WC}).
This is reminiscent of the difference observed in the context of $\Lambda$CDM, where EC priors also lead to lower $A_s$ and $n_s$ values \cite{Simon:2022lde,Holm:2023laa}.
Several potential sources for the difference in the profiles (of \cref{fig:profile_EFT}) are discussed in \cref{app:EC-vs-WC}. Our test indicates that the different EFT priors primarily explain this (as seen in the context of $\Lambda$CDM \cite{Simon:2022lde,Holm:2023laa}), and therefore, any preference (or lack thereof) in the EFTofBOSS analysis should be seen as prior-dependent (since the priors have been promoted to effective likelihoods in the Frequentist analysis). We delve deeper into this issue in \cref{app:EC-vs-WC}. Fortunately, the effect of the EFT priors becomes less important when we combine EFTofBOSS with CMB data, and we anticipate that the profile likelihood will not be strongly affected by the choice of EFT priors.

\indent We also conduct both Bayesian (reported in \cref{table:BOSS+eBOSS parameters}) and frequentist analyses when incorporating EFTofeBOSS QSO data to assess whether these data can improve the agreement with CMB data or further constrain the SI$\nu$ mode. As evident in \cref{fig:profile_EFT}, the inclusion of eBOSS QSO data does not significantly alter the profile likelihood compared to that obtained with EFTofBOSS+BBN data, and the disagreement with CMB data only worsens (slightly).

\subsection{Combined CMB and LSS constraints}

\renewcommand{\arraystretch}{1.5}
\begin{table*}
 \caption{68\% CL intervals for cosmological parameters (or 95\% for upper limit constraints) obtained when analyzing TTTEEE CMB data from \textsc{CamSpec NPIPE} combined with the EFTofBOSS+BBN dataset for the two self-interacting neutrino modes as well as with added Ly-$\alpha$ for SI$\nu$. We also display the $\Delta\chi^2$ with respect to $\Lambda$CDM, as well as the frequentist confidence interval for $\log_{10}{G_{\rm eff, \nu}/{\rm MeV^{-2}}}$.}
\label{table:BOSS+NPIPE parameters}
\begin{tabular}{|c|c|c|c|}
\hline
\multicolumn{4}{|c|}{\sc{{\it Planck} TTTEEE+EFTofBOSS}} \\
\hline
\hline
& MI$\nu$ & SI$\nu$ & SI$\nu$ (+Ly-$\alpha$)\\
\hline
$\rm log_{10} G_{\rm eff, \nu} $
	 & $<$ -3.52(-5.0)
	 & -1.72(-1.73)$ ^{+0.17}_{-0.089}$ 
     & -1.687(-1.683)$ \pm 0.034$ 
	 \\
$-2\log{\cal L}(\log_{10} G_{\rm eff, \nu})$ & $<$ -3.36  & -1.73$^{0.12}_{-0.14}$ & $-$  \\  
\hline
$H_0$
	 & 66.9(67.1)$ \pm 1.1$ 
	 & 67.5(67.9)$ \pm 1.2$  
& 67.90(68.21)$ \pm 0.95$ 
	 \\
$10^{-2}\omega{}_{b }$
	 & 2.214(2.212)$ \pm 0.018$ 
	 & 2.226(2.225)$ \pm 0.018$  
     &  2.230(2.228)$ \pm 0.016$
	 \\
$\omega{}_{\rm cdm }$
	 & 0.1175(0.1173)$ \pm 0.0022$ 
	 & 0.1172(0.1167)$ \pm 0.0025$  
     & 0.1173(0.1175)$ \pm 0.0020$ 
	 \\
$n_{s }$
	 & 0.9586(0.9588)$ \pm 0.0069$ 
	 & 0.9317(0.9312)$ \pm 0.0058$  
     & 0.9334(0.9323)$ \pm 0.0048$
	 \\ 
$10^{9}A_{s }$
	 & 2.081(2.077)$ \pm 0.035$ 
	 & 1.964(1.952)$ \pm 0.032$  
     & 1.968(1.961)$ \pm 0.029$ 
	 \\
$\tau{}_{\rm reio }$
	 & 0.0545(0.0532)$ \pm 0.0077$ 
	 & 0.0547(0.0525)$ \pm 0.0074$ 
     & 0.0564(0.0543)$ ^{+0.0063}_{-0.0073}$ 
 \\
$\sum m_\nu$/eV
	 & $<$ 0.135(0.0005) 
	 & $<$ 0.192(0.0005)  
     & $<$ 0.125 (0.0004)
	 \\
$\Delta N_{\rm{eff}}$
	 & $-0.12(-0.14) ^{+0.13}_{-0.15}$ 
	 & $-0.19(-0.22) \pm 0.15$  
     & -0.18(-0.18)$ \pm 0.12$ 
	 \\
$S_8$
	 & 0.824(0.827)$ \pm 0.011$ 
	 & 0.817(0.824)$ \pm 0.011$  
     & 0.820(0.824)$ \pm 0.010$ 
	 \\
\hline
 $\Delta \chi^2 =\chi^2_{\rm min}-\chi^2_{{\rm min},\Lambda \rm CDM}$  & -$3.27$ & +1.25  & -19.83\\
 \hline

\end{tabular}
\end{table*}

\begin{figure*}
    \centering
    \includegraphics[width=1.48\columnwidth]
    {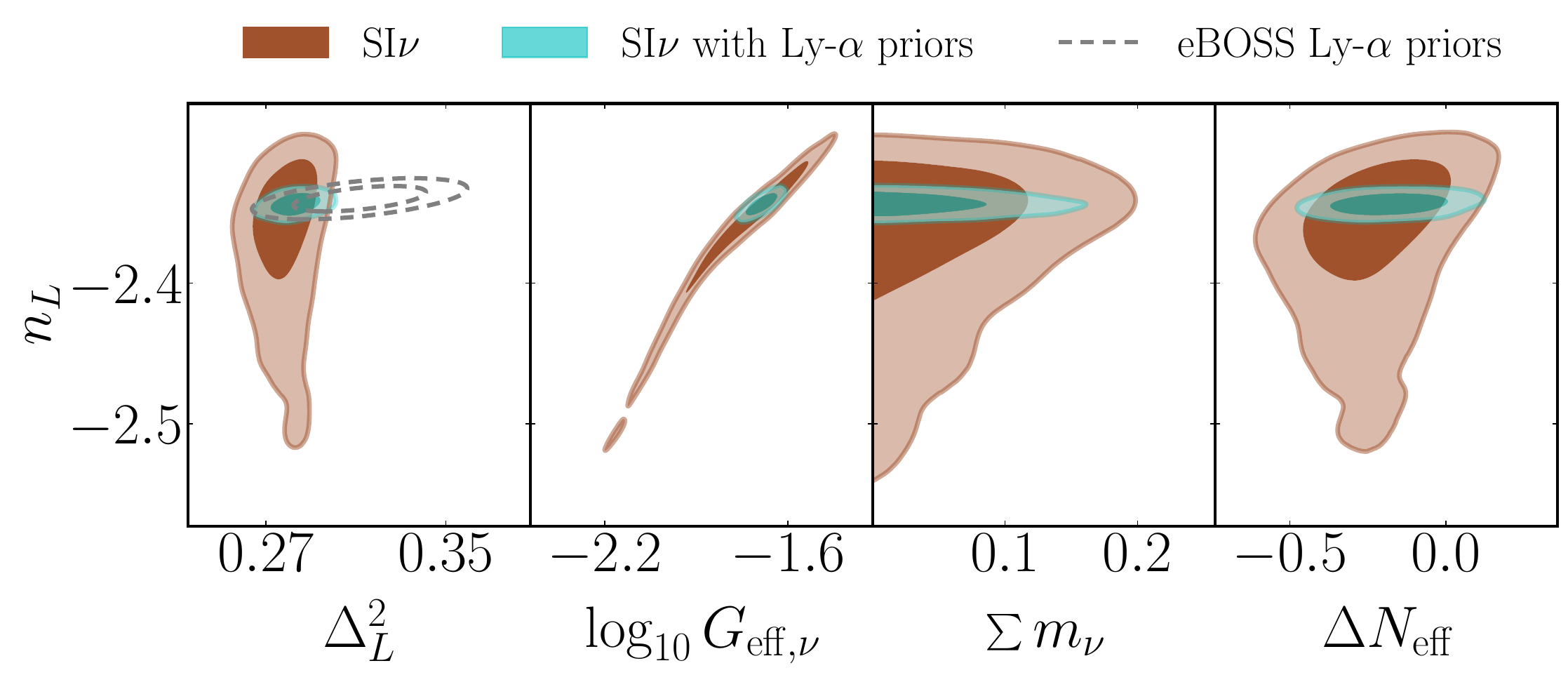}
    \includegraphics[width=0.52\columnwidth]{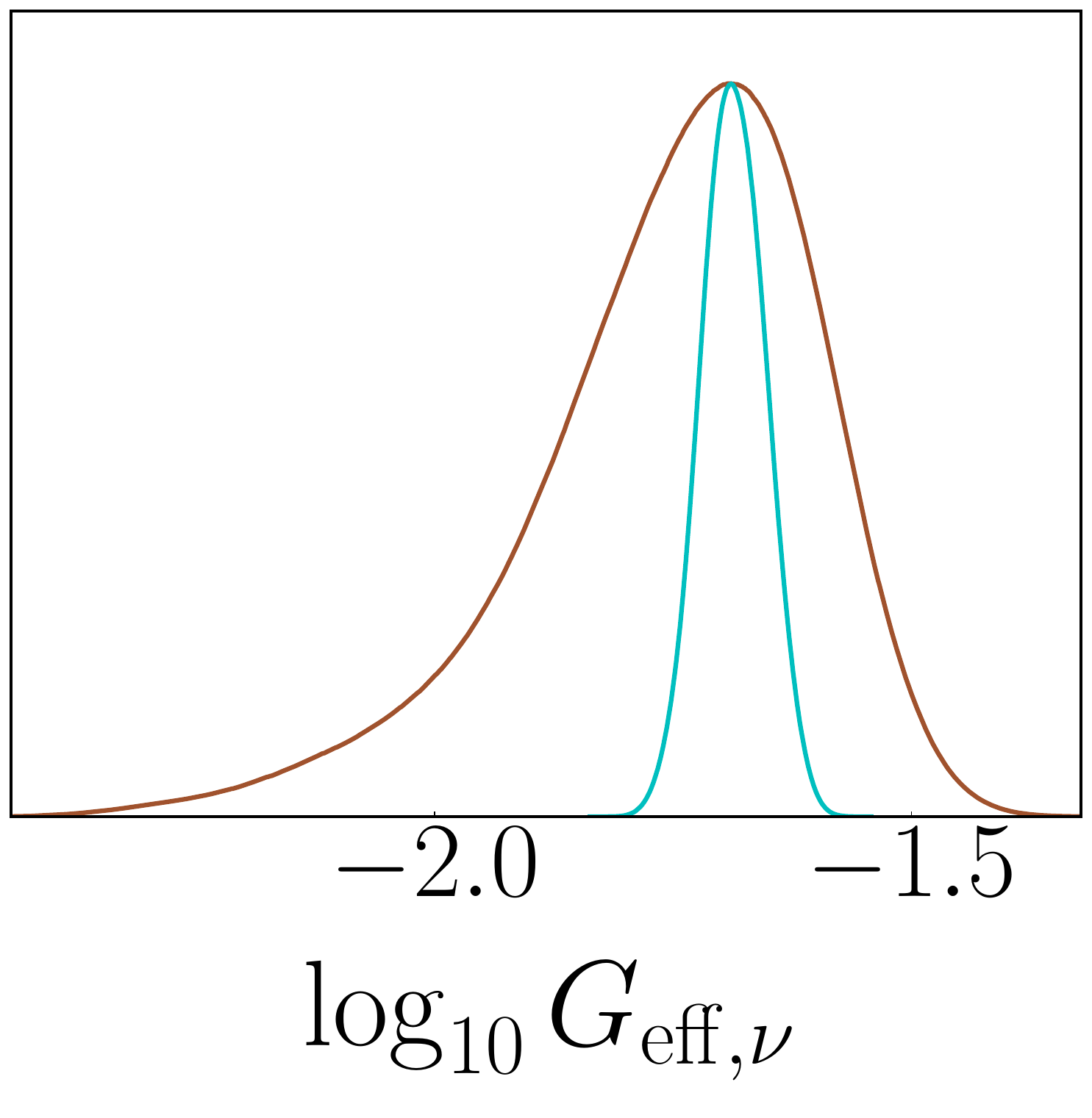}
    \caption{2-dimensional posteriors of $n_L$ for the SI$\nu$ mode constrained by \textsc{CamSpec NPIPE} + EFTofBOSS + EFTofeBOSS + BBN, with and without the Ly-$\alpha$ prior on eBOSS ({\it Left}). And 1-dimensional posterior of \logGeff for the same datasets ({\it Right}).}
        \label{fig:2-d_Lyz}
\end{figure*}

We now perform a combined analysis of \textsc{CamSpec NPIPE} and EFTofBOSS + BBN data. 
We present our results in \cref{table:BOSS+NPIPE parameters} and plot the posterior distributions in Fig.~\ref{fig:2-d_EFT} (empty contours), and the profile likelihood in \cref{fig:profile_EFT} (right panel) alongside those obtained from each dataset separately.
When combining CMB and EFTofBOSS data, we find a preference for the MI$\nu$ mode over the SI$\nu$ mode, with $\Delta \chi^2$(SI$\nu$-MI$\nu$) $=-4.52$. 
Since the EFTofBOSS + BBN and \textsc{CamSpec NPIPE} data disagree on the location of the SI$\nu$ mode, it becomes disfavored when combining the data (with $\Delta\chi^2({\rm SI\nu})= + 1.25$ compared to $\Lambda$CDM).
This is in good agreement with Ref.~\cite{Camarena_2025}, which used the \texttt{CLASS-PT} code that incorporates the EC priors  (see App.~\ref{app:EC-vs-WC} for mode details on differences), showing a good agreement between the EC and WC analyses of BOSS when combined with \textit{Planck}.
In addition, it is important to note that $\Lambda$CDM remains favored over the MI$\nu$ mode when considering that the small improvement in $\chi^2$ comes at the cost of three additional free parameters.

\subsection{Impact of Ly-$\alpha$ data on the SI$\nu$ mode}

Finally, we include eBOSS 1D Ly-$\alpha$ flux power spectrum data in the analysis, as was previously done in Ref.~\cite{He:2023oke}. Note that these data are in significant tension with $\Lambda$CDM \cite{Rogers:2023upm}. We limit our analysis to the SI$\nu$ mode, as the MI$\nu$ mode cannot explain the tension.

In \cref{fig:2-d_Lyz}, we present the posterior distribution of $n_L$ against $\Delta_L^2$ and the neutrino model parameters -- namely \logGeff, $m_\nu$, and $\Delta N_{\rm eff}$ -- for the \textsc{CamSpec NPIPE} + EFTofBOSS/eBOSS + BBN, with and without the Ly-$\alpha$ prior on eBOSS. 
We also show the 1D posterior distribution of \logGeff. In \cref{table:BOSS+NPIPE parameters}, we display the cosmological constraints, while in \cref{table:contributions}, we provide the contribution of each dataset to the $\Delta\chi^2$ relative to $\Lambda$CDM.
The $n_L$-vs-$\Delta_L^2$ plane demonstrates the excellent agreement between the analyses with and without the eBOSS Ly-$\alpha$ prior for the SI$\nu$ mode, implying that the strongly-interacting scenario could resolve the Ly-$\alpha$ tension present in $\Lambda$CDM. 
As a result, incorporating the Ly-$\alpha$ data leads to a strong preference for the SI$\nu$ mode $(\Delta\chi^2 = -19.83)$, primarily driven by a better fit to the Ly-$\alpha$ eBOSS priors and {\sc CamSpec NPIPE TTTEEE}, as evident in \cref{table:contributions}, and a very tight determination of \logGeff. However, we find a slight degradation in the EFTofBOSS data that was not found in Ref.~\cite{He:2023oke}. This slight discrepancy is likely attributed to the difference in the treatment of priors between the EC and WC datasets.  

\renewcommand{\arraystretch}{1.5}
\begin{table}[h!]
 \caption{Individual contributions to $\Delta \chi^2$ for the strongly-interacting mode analysis with {\textsc{CamSpec NPIPE}} TTTEEE + EFTofBOSS/eBOSS + BBN + Ly-$\alpha$.}
\label{table:contributions}
\begin{tabular}{|c|c|}
\hline
& $\Delta \chi^2 =\chi^2_{\rm min,SI \nu}-\chi^2_{{\rm min},\Lambda \rm CDM}$\\
\hline
{\sc{{\it Planck}}} low-$\ell$ TT & $+0.71$\\
{\sc{{\it Planck}}} low-$\ell$ EE &  $+0.16$\\
{\sc{{\it Planck}}} lensing & $ -0.3$\\
{\textsc{CamSpec NPIPE}} TTTEEE &  $-6.23$\\
EFTofBOSS & $+4.03$ \\
EFTofeBOSS &  $+0.01$\\
\textsc{BBN} & $-2.24$\\
Ly-$\alpha$ eBOSS priors &  $-15.97$\\
\hline
Total &  $-19.83$\\
\hline
\end{tabular}
\end{table}
\section{Conclusions}
\label{conclusion}

Neutrinos are ubiquitous in cosmology, leaving their imprints on our cosmological probes across multiple epochs \cite{Lesgourgues:2014zoa}, hereby giving us the opportunity to learn about their properties with high precision observation beyond the mere value of their absolute mass scale. 
Exotic neutrino properties, such as self-scattering, have garnered significant attention in the context of the Hubble and $S_8$ tensions.

Indeed, as reviewed in \cref{model}, it has been found that the {\it Planck} temperature data described with the \textsc{Plik} 2018 likelihood allow for two distinct interaction modes in the posterior distribution: one mode corresponding to moderately interacting neutrinos, aligning with the standard model expectations and the $\Lambda$CDM values for cosmological parameters; and another mode of strongly-interacting neutrinos, resulting in higher $H_0$ and lower $S_8$ values \cite{Kreisch:2019yzn}.
ACT DR4 data further support neutrino strong self-interactions, for both temperature and polarization \cite{Kreisch:2022zxp}. 
Interestingly, recent findings show that BOSS large-scale structures (LSS) data \cite{Camarena_2023} and Lyman-$\alpha$ data \cite{He:2023oke} also favor this mode.  
However, the {\it Planck} polarization data appear to disfavor that possibility \cite{Kreisch:2019yzn,Camarena_2025}.

Recently, a new set of CMB maps called  NPIPE  have been released by the {\it Planck} collaboration \cite{Planck:2020olo}, and new likelihoods have been subsequently built out of these maps \cite{Rosenberg:2022sdy,Tristram:2023haj}.  
Moreover, new sets of BAO measurements by the DESI collaboration have been released.
This called for an update in the analysis of the strongly-interacting neutrino model, in order to check whether the new {\it Planck} and DESI data would further constrain the model, or instead align with the ACT and LSS datasets. 

In this work, we have updated the constraints on the self-interacting neutrino model in light of the improved {\it Planck} data, modeled using the {\sc CamSpec} likelihood \cite{Rosenberg:2022sdy}. Following Refs.~\cite{Camarena_2023,Camarena_2025}, we have compared results from both Bayesian and Frequentist approaches. 
We have also considered the impact of BAO data, contrasting results obtained with BOSS and DESI.
Furthermore, we have re-evaluated constraints from LSS data using an alternative implementation of EFTofBOSS, thanks to the \texttt{PyBird} code \cite{DAmico:2020kxu}, and included, for the first time, the eBOSS quasar data alongside the previously used galaxy data.
Our results can be summarized as follows:

\begin{enumerate}
    \item[\textbullet] {\bf Plik18 vs NPIPE:} While {\sc Plik} 2018 significantly favored the MI$\nu$ mode, the preference has diminished with the new {\sc CamSpec} data. The SI$\nu$ mode no longer appears significantly disfavored, and the constraints to $G_{\rm eff,\nu}$ in the MI$\nu$ mode have become slightly weaker. This may seem counter-intuitive, as one might expect more accurate data to further constrain the SI$\nu$ mode. However, it demonstrates that {\sc NPIPE} data do not rule out the possibility of strongly-interacting neutrinos. 
     \item[\textbullet] {\bf Cosmic tensions:} Nevertheless, the possibility of SI$\nu$ resolving the Hubble tension remains excluded. The only cosmological parameters that appear strongly affected (by $1-2\sigma$) in the SI$\nu$ mode are $A_s$ and $n_s$, as shown in \cref{fig:2-d_NPIPE} for both {\sc Plik} 2018 and {\sc CamSpec NPIPE} data. This should be seen as a limitation to the robustness of the constraints to $A_s$ and $n_s$ in the $\Lambda$CDM model.
     \item[\textbullet] {\bf BOSS vs DESI:} While the inclusion of SDSS BAO (and Pantheon$^+$) data does not alter these conclusions, the new DESI BAO data shift the preference towards the SI$\nu$ mode. The constraints to $\sum m_\nu$ also significantly relax compared to the standard DESI result, with $\sum m_\nu < 0.093$ eV (MI$\nu$ mode) and $\sum m_\nu<0.18$ eV (SI$\nu$ mode).
     \item[\textbullet]{\bf `EC vs WC' EFTofLSS:} While we also find two interaction modes when analyzing BOSS data through the EFTofLSS under the WC parametrization (with \texttt{PyBird}), neither mode is favored over the other. This contrasts with results obtained using the EC parametrization of the EFTofLSS, which suggests a slight preference for the SI$\nu$ mode. We trace the bulk of this difference to the different priors set on EFT parameters, as discussed in \cref{app:EC-vs-WC}. The inclusion of eBOSS QSO data does not significantly alter the results.
     \item[\textbullet]{\bf EFTofBOSS vs NPIPE:} The region of interest for the interaction in the SI$\nu$ mode differs between the CMB and EFTofBOSS analyses. Consequently, the combined analysis of \textsc{CamSpec NPIPE} and EFTofBOSS favors the MI$\nu$ mode over SI$\nu$. However, the preference for any mode of interaction over $\Lambda$CDM is minimal. Considering the difference between (e)BOSS full-shape and DESI BAO data, it would be interesting to conduct this analysis using DESI full-shape data (currently not publicly available).
     \item[\textbullet]{\bf Lyman-$\alpha$:} When combining with eBOSS Ly-$\alpha$, we find a strong preference for the SI$\nu$ mode, even though the improvement in the fit is slightly reduced ($\Delta\chi^2\sim-20$) when using WC rather than the EC implementation of the EFTofLSS \cite{He:2023oke}.

\end{enumerate}

In conclusion, while neither CMB nor LSS data alone conclusively rule out the possibility of strongly-interacting neutrinos, the combination of both appears to favor the MI$\nu$ mode over the SI$\nu$ mode, regardless of the EFTofLSS implementation (EC vs WC) or version of {\it Planck} data ({\sc Plik 2018} or {\sc CamSpec NPIPE}). However, the results of the combined {\it Planck}+EFTofBOSS analysis seem to contradict the results of the {\it Planck}+DESI BAO analysis, which favors the SI$\nu$ mode. Therefore, the DESI data analyzed under the EFTofLSS are particularly anticipated to shed light on this disagreement. It would also be interesting to consider updated analyses of eBOSS Ly-$\alpha$ data \cite{Fernandez:2023grg,Walther:2024tcj} to test the preference observed here and elsewhere \cite{He:2023oke}.
Finally, we acknowledge that our analyses are limited to the simple model of universal self-interacting neutrinos for each flavor, following a four-fermion interaction. Different conclusions may be reached for other, perhaps more realistic, models of self-interacting neutrinos \cite{Blinov:2019gcj}. We leave this exploration for future work.

\textbf{Acknowledgments} -- 
 This work is supported by funding from the European Research Council (ERC) under the European Union's HORIZON-ERC-2022 (grant agreement no.~101076865) and from the European Union's Horizon 2020 research and innovation program under the Marie Sk{\l}odowska-Curie grant agreement no.~860881-HIDDeN. The authors acknowledge the use of computational resources from the LUPM's cloud computing infrastructure founded by Ocevu labex and France-Grilles. 

\newpage
\appendix
\newpage
\section{Comparison with the ``East Coast'' analysis}
\label{app:EC-vs-WC}

\renewcommand{\arraystretch}{1.5}
\begin{table*}
 \caption{68\% CL intervals for cosmological parameters (or 95\% for upper limit constraints) obtained when analyzing EFTofBOSS+BBN dataset using the West Coast (WC) and East Coast (EC) priors for the two self-interacting neutrinos modes. We also display the $\chi^2$ comparison between both modes and $\Lambda$CDM.}
\label{table:BOSSWC_BOSSEC parameters}
\begin{tabular}{|c|c|c|c|c|}
\hline
&\multicolumn{2}{c}{MI$\nu$} & \multicolumn{2}{|c|}{SI$\nu$}\\
\hline
\hline
& EFTofBOSS WC  &EFTofBOSS EC & EFTofBOSS WC  &EFTofBOSS EC\\
\hline
$\rm log_{10} G_{\rm eff, \nu} $
	 & $<$$ -2.40(-3.13)$
	 & $<$ $-2.14(-2.0) $
	 & $-0.98(-1.06)$$ ^{+0.39}_{-0.32}$ 
	 & $-1.19(-1.29)$$ ^{+0.32}_{-0.22}$ 
	 \\
$H_0$
	 & 68.4(67.7)$ ^{+1.8}_{-2.1}$ 
	 & 68.9(67.9)$ ^{+1.8}_{-2.3}$ 
	 & 68.7(67.8)$ ^{+1.7}_{-2.2}$ 
	 & 68.9(67.8)$ ^{+1.9}_{-2.1}$ 
	 \\
$10^{-2}\omega{}_{\rm b }$
	 & 2.260(2.245)$ \pm 0.060$ 
	 & 2.253(2.254)$ \pm 0.061$ 
	 & 2.254(2.247)$ \pm 0.059$ 
	 & 2.253(2.238)$ \pm 0.061$ 
	 \\
$\omega{}_{\rm cdm }$
	 & 0.1248(0.1178)$ ^{+0.0095}_{-0.013}$ 
	 & 0.1387(0.1294)$ ^{+0.0099}_{-0.015}$ 
	 & 0.1218(0.1127)$ ^{+0.0085}_{-0.014}$ 
	 & 0.136(0.127)$ ^{+0.011}_{-0.014}$ 
	 \\
$n_{s }$
	 & 0.957(0.935)$ ^{+0.078}_{-0.087}$ 
	 & 0.906(0.963)$ \pm 0.080$ 
	 & 0.946(0.959)$ ^{+0.090}_{-0.080}$ 
	 & 0.852(0.879)$ \pm 0.087$ 
	 \\
$10^{9}A_{s }$
	 & 2.20(2.13)$ ^{+0.35}_{-0.46}$ 
	 & 1.43(1.76)$ ^{+0.20}_{-0.29}$ 
	 & 2.08(2.4)$ ^{+0.35}_{-0.42}$ 
	 & 1.37(1.54)$ ^{+0.17}_{-0.30}$ 
	 \\
$\sum m_{\nu }$/eVs
  & $<$ 1.06(0.0003
	 & $<$ 0.844(0.21) 
	 & $<$ 0.926(0.26)
	 & $<$ 0.917(0.13)
	 \\
$\Delta N_{\rm{eff}}$
	 & $-0.05(-0.12)$$ ^{+0.28}_{-0.25}$ 
	 & $-0.08(-0.08)$$ ^{+0.25}_{-0.30}$ 
	 & $-0.098(-0.119)$$ \pm 0.25$ 
	 & $-0.09(-0.16)$$ \pm 0.27$ 
	 \\
$S_8$
	 & 0.819(0.838)$ \pm 0.052$ 
	 & 0.736(0.835)$ \pm 0.047$ 
	 & 0.787(0.812)$ ^{+0.046}_{-0.052}$ 
	 & 0.695(0.747)$ ^{+0.043}_{-0.054}$ 
	 \\
\hline
 $\Delta \chi^2 =\chi^2_{\rm best-fit}-\chi^2_{\Lambda \rm CDM}$ & $-3.12 $&$ -0.90$  &$-1.07$ & $-3.32 $\\
 \hline
\end{tabular}
\end{table*}

As mentioned in the main text, our profile appears to be in (slight) disagreement with that derived in previous works \cite{Camarena_2023,Camarena_2025}. 
In contrast to previous self-interacting neutrino LSS analyses~\cite{He:2023oke,Camarena_2023,Camarena_2025}, we consider several different analysis choices in this work.\footnote{For instance, we use different EFT priors (all previous analyses used the EC priors), different cutoff scales $k_{\rm max}$, and we do not consider the $Q_0$ metric~\cite{Ivanov:2021fbu} (or equivalently the $\slashed{P}$ metric~\cite{DAmico:2021ymi}). We also do not include the bispectrum monopole (as Ref.~\cite{He:2023oke}), we use slightly different data (obtained with different estimators), and we use a different code (all previous analyses used the \texttt{CLASS-PT} code) with different implementations, leading to minor variations (of the IR resummation for example). All these differences are explained in detail in Ref.~\cite{Simon:2022lde}.}
According to Ref.~\cite{Simon:2022lde}, the main difference between the $\Lambda$CDM constraints derived from the \texttt{PyBird} likelihood and those derived from the \texttt{CLASS-PT} likelihood lies in the priors on the EFT parameters. 
It is now well established that EFTofBOSS analyses suffer from prior dependence and prior volume effects: analyses with different priors on the EFT parameters (especially those from the EC and the WC baselines), can lead to results that differ at the $1\sigma$ level on the $\Lambda$CDM parameters (in particular on $A_s$, $\sigma_8$ and $\Omega_m$)~\cite{Simon:2022lde,Carrilho:2022mon,Holm:2023laa,Gsponer:2023wpm,Zhao:2023ebp,Nishimichi:2020tvu}. 
To investigate this difference in the self-interacting neutrino case, we performed, with the \texttt{PyBird} code, the EFTofBOSS (+ BBN) analysis with the EC priors, as presented in \cref{table:BOSSWC_BOSSEC parameters}.
Our results can be summarized as follows:  
\begin{itemize}
    \item the $68\%$ CL constraint on $\Delta N_{\rm eff}$ remains stable under this change of prior at $\lesssim 0.1 \sigma$;
    \item the $95\%$ CL constraint on $\sum m_{\nu}$ remains stable for the SI$\nu$ case, while we obtain a $30\%$ better constraint with the EC priors compared to the WC priors for the MI$\nu$ case;
    \item the $95\%$ CL constraint on $\rm log_{10} G_{\rm eff, \nu} $ improves by $10 \%$ with the EC priors in the MI$\nu$ case, while we obtain a shift of $0.7 \sigma$ for the $68\%$ CL constraint in the SI$\nu$ case. 
\end{itemize}
Our constraints with the EC priors appear in good agreement with those presented in Ref.~\cite{Camarena_2023}, attesting that prior choices are indeed the cause of the discrepancy between our analyses and those presented therein.
Although the EFT prior effects are not negligible, they are within the expected range based on previous analyses in the $\Lambda$CDM context. When combining EFTofBOSS with \textit{Planck}, we expect the effect of the EFT priors on the posteriors to become negligible.

\begin{figure}[h!]
    \centering
    \includegraphics[width=\columnwidth]{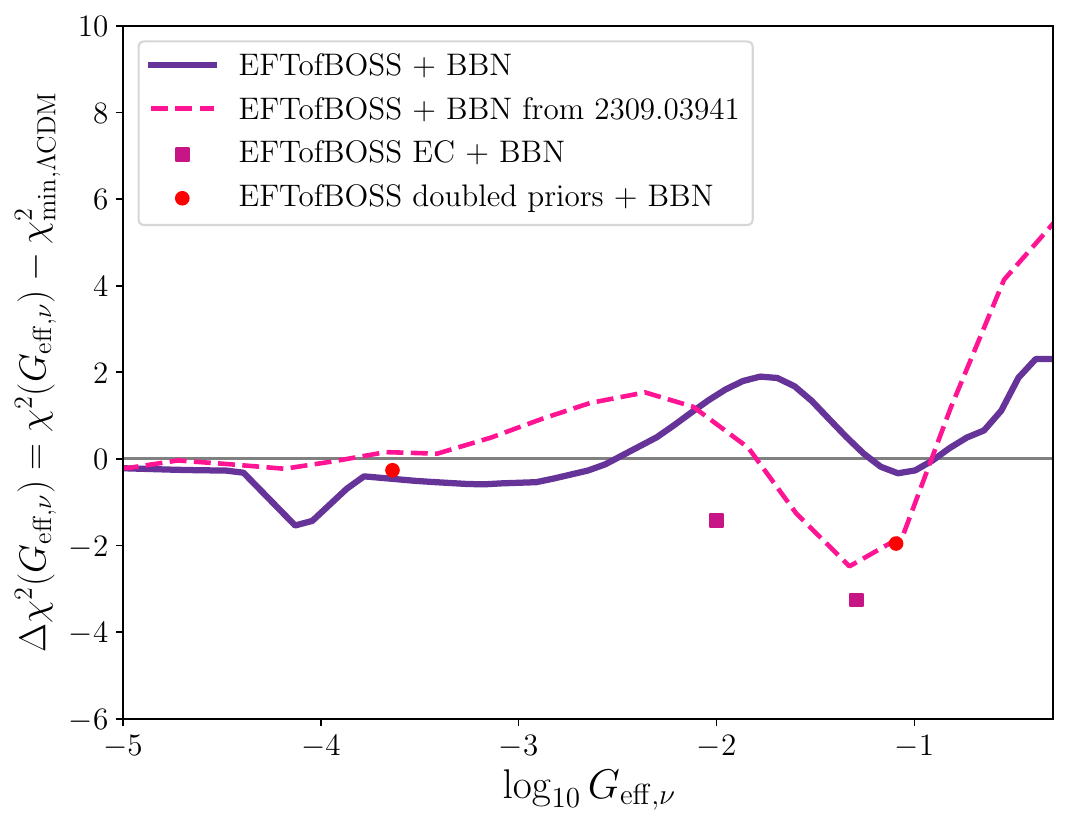}
    \caption{Profile likelihood (normalized to $\Lambda$CDM) of $G_{\rm eff,\nu}$ for EFTofBOSS + BBN from our analysis and from Ref.~\cite{Camarena_2023}, together with the best-fits of the two modes from the ``East Coast'' priors (pink rectangles) and from the ``West Coast'' priors multiplied by two (red circles).}
    \label{fig:EC_vs_WC}
\end{figure}

To explore the impact of the EFT priors (when promoted to effective likelihoods) on the profile likelihood, \cref{fig:EC_vs_WC} presents the best-fits of the two modes derived from the EC priors (pink rectangles) and those derived from the WC priors multiplied by two (red circles), alongside our EFTofBOSS profile and the one from Ref.~\cite{Camarena_2023}. Notably, we are able to replicate the SI$\nu$ best-fit of Ref.~\cite{Camarena_2023} with our EC analysis,\footnote{This is not the case for the MI$\nu$ best-fit, likely due to the flat $\chi^2$ region.} which reinforces that the bulk of the difference between our baseline analysis and theirs stems from the different choice of EFT priors. 

Furthermore, we conduct an analysis of the WC priors, where we double their size. In that case, \cref{fig:EC_vs_WC} reveals that we can approximately replicate the best-fit values of Ref.~\cite{Camarena_2023}. 
As explained in Refs.~\cite{Simon:2022lde,Holm:2023laa}, the WC priors are more informative than the EC priors.
However, expanding the priors comes at the cost of increasing the prior volume projection effect in the posteriors and potentially allowing EFT parameters to take on non-physical values (possibly breaking the perturbative nature of the EFTofLSS).
Lastly, let us emphasize that we expect this prior dependence to decrease significantly when we combine EFTofBOSS with \textit{Planck}, as evidenced by the good agreement between our analysis and Ref.~\cite{Camarena_2025} (which used the \texttt{CLASS-PT} code that incorporates the EC prior).
\clearpage
\bibliographystyle{apsrev4-1}
\bibliography{Bibliography}
\end{document}